\PassOptionsToPackage{table}{xcolor}
\documentclass[letterpaper,twocolumn,10pt]{article}
\usepackage{hegroup}
\usepackage[table]{xcolor}
\usepackage[normalem]{ulem}
\useunder{\uline}{\ul}{}
\usepackage{tikz}
\usepackage{amsfonts}
\usepackage{xspace} 
\usepackage{amsmath}
\usepackage{mathrsfs}
\usepackage{amssymb}
\usepackage{enumitem}

\usepackage{amsthm}
\usepackage{multirow}
\usepackage{makecell}
\usepackage{subcaption}
\usepackage{float}
\usepackage{booktabs}
\usepackage{balance}
\usepackage{tcolorbox}
\usepackage{xurl}
\usepackage{hyperref}
\usepackage{tabularx}
\usepackage{cleveref}
\usepackage{algorithm}
\usepackage{algpseudocode}
\newtheorem{definition}{Definition}

\usepackage{footmisc}

\newcommand{\mypara}[1]{\noindent{\bf {#1}.}}

\begin{document}

\title{Auditing Data Membership in Reinforcement Learning \\With Verifiable Rewards}
\date{}

\author{
Yule Liu\textsuperscript{1}    \ \ \
Heyi Zhang \textsuperscript{2}  \ \ \
Jingyi Zheng\textsuperscript{1} \ \ \
Zhen Sun\textsuperscript{1} \ \ \
Zifan Peng\textsuperscript{1} \ \ \ \\
Jiaheng Wei\textsuperscript{1} \ \ \
Tianshuo Cong\textsuperscript{3} \ \ \
Yilong Yang\textsuperscript{4}\textsuperscript{\textdagger} \ \ \
Xinlei He\textsuperscript{5}\textsuperscript{\textdagger} \ \ \
\\
\textsuperscript{1}\textit{The Hong Kong University of Science and Technology (Guangzhou)} \ \ \  \\
\textsuperscript{2}\textit{Shanghai Jiao Tong University} \ \ \ 
\textsuperscript{3}\textit{Shandong University} \ \ \ 
\textsuperscript{4}\textit{Xidian University} \ \ \ 
\textsuperscript{5}\textit{Wuhan University} \ \ \ 
}

\maketitle
\footnotetext{\textsuperscript{\textdagger}Co-corresponding authors: Yilong Yang (\href{mailto:yilongyang@xidian.edu.cn}{yangyilong@xidian.edu.cn
}) and Xinlei He (\href{mailto:xinleihe@hkust-gz.edu.cn}{xinlei.he@whu.edu.cn}).}

\begin{abstract}
Reinforcement Learning with Verifiable Rewards (RLVR) has become a core training stage in recent large language models (LLMs). 
Its reliance on non-public, high-value prompt sets raises concerns about unauthorized data use, creating a need for exposure auditing.
A natural tool is membership inference attacks (MIAs), but existing methods detect fitting to a fixed target string. 
This does not apply to RLVR, which generates responses from the model itself and reinforces successful ones, thus hindering the auditing of data exposure.
We show that it remains detectable: RLVR reshapes the model's response distribution on training prompts, producing behavioral traces that can be surfaced through targeted auditing.
  
We propose Divergence-in-Behavior Auditing (DIBA), a white-box query-level auditing framework for RLVR.
DIBA compares a fine-tuned model against its pre-RLVR checkpoint along two axes: reward-side evidence capturing changes in verifiable task success, and policy-side evidence capturing prompt-conditioned behavioral drift.
By aggregating over multiple stochastic rollouts, DIBA produces a stable query-level auditing signal.

Under a white-box setting, DIBA consistently outperforms strong transferred likelihood-based baselines, including calibrated and self-generated variants, achieving around 0.8 AUC and an order-of-magnitude stronger TPR@0.1\%FPR.
We further show that RLVR auditing is stronger when training leaves non-trivial prompt-specific traces and weaker when the base model already performs well on the prompt. 
Under a practical grey-box setting, transfer is often robust across model sizes under the same RLVR algorithm, but more varied across algorithms, and can remain useful under distribution shift with carefully chosen shadow data. 
We also extend the analysis to vision-language models and study possible mitigations, finding that post-hoc output transformations can reduce text-level auditability while standard training regularization provides limited protection effects. 
Overall, our results show that exposure in RLVR is better audited through prompt-conditioned behavioral traces than fixed-target fitting, and position DIBA as a practical auditor for white-box RLVR compliance settings.
\end{abstract}

\section{Introduction}
Reinforcement Learning with Verifiable Rewards (RLVR)~\cite{shao2024deepseekmath,wen2025reinforcement,yu2025dapo,DBLP:journals/corr/abs-2507-04136,he2025sample} has become a core training stage for improving reasoning and other objectively verifiable capabilities in recent LLMs, e.g., DeepSeek-R1/V4~\cite{shao2024deepseekmath}, Kimi-k2.5~\cite{team2025kimi}, and GLM-5~\cite{zeng2026glm}.
Its effectiveness depends heavily on high-quality prompts with static verifiers, which are often non-public and commercially valuable, such as proprietary benchmarks~\cite{dong2024generalization} and private code or agentic pipelines~\cite{jimenez2023swe}.
This creates a practical need for data exposure auditing: owners of such prompt suites may wish to determine whether their data is used in a released RLVR pipeline without authorization.

\begin{figure}[t]
    \centering
    \includegraphics[width=\linewidth]{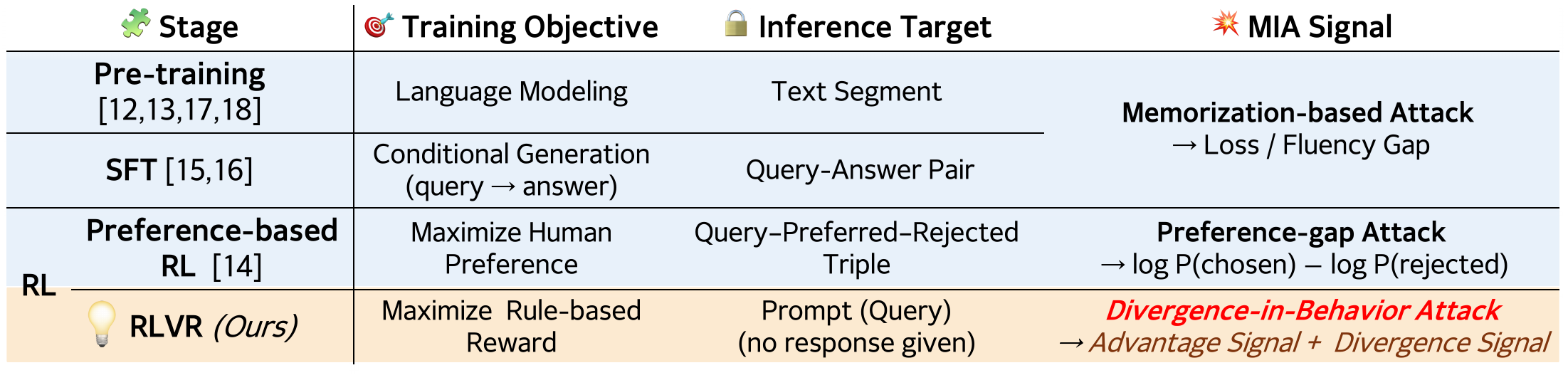}
    \caption{MIAs in different LLM training paradigms.}
    \label{fig:rlmia}
\end{figure}

Membership inference attacks (MIAs) provide a standard lens for auditing training-data exposure in LLMs. 
Prior work has studied membership risks in several stages of the LLM training pipeline, including pre-training~\cite{xie2024recall,zhang2024min}, supervised fine-tuning (SFT)~\cite{ran2025lora,fu2023practical}, and preference-based alignment~\cite{feng2024exposing}.
As shown in \Cref{fig:rlmia}, although these attacks operate on different granularities of inference targets, e.g., a text segment (pre-training), a query-response pair (SFT), or a preference tuple (preference RL), they all audit fitting to a fixed, known target string.
RLVR breaks this assumption: rather than optimizing toward a predetermined response, it generates candidate responses from the model itself, reinforces successful ones, and inhibits the failed ones~\cite{shao2024deepseekmath,yu2025dapo}.
The shift in training algorithm motivates a new formulation, i.e., \emph{given a query and model weights, the auditor aims to determine whether it was used during RLVR post-training.}

Adapting existing MIAs to this setting is non-trivial, as no fixed target response exists to score against.
To systematically understand the data exposure risk in RLVR, we consider a commonly used white-box auditing scenario for LLMs~\cite{fu2023practical,ran2025lora,xie2024recall} and assume access to two types of knowledge. 
First, the auditor can obtain the model's pre-RLVR checkpoint, i.e., the weights before RLVR training is applied.
This is realistic when the deployed model family requires disclosing its foundation model used for fine-tuning~\cite{llama-license} or when the checkpoint can be matched through model fingerprinting~\cite{pasquini2025llmmap}. 
Second, for many verifiable domains such as math reasoning, the verifier is rule-based and accessible, or can be faithfully reimplemented by the auditor~\cite{numina_math_datasets}.

\mypara{Auditing Approach}
Our key observation is that, although RLVR does not optimize toward a fixed target response, it can still leave {observable post-training traces} when the fine-tuned policy is compared against its pre-RLVR reference policy.
These traces arise along two complementary dimensions.
At the task level, RLVR may change how often the model achieves verifiable success on a prompt.
At the policy level, RLVR may alter how probability mass is distributed over generated trajectories.
Based on these insights, we propose \textbf{D}ivergence-\textbf{i}n-\textbf{B}ehavior \textbf{A}uditing (DIBA), a query-level auditing framework for RLVR. 
Concretely, DIBA compares the two policies along two complementary axes:
\begin{itemize}[leftmargin=*]
    \item \textit{Reward-side evidence}, which captures whether RLVR changes verifiable task success on a prompt and contributes to the overall ranking capability;
    \item \textit{Policy-side evidence}, which captures whether RLVR induces behavioral drift relative to the reference checkpoint and contributes to the high-precision signals for low-FPR detection.
\end{itemize}
By averaging over multiple stochastic rollouts from the fine-tuned and reference policies, DIBA estimates a stable query-level auditing signal under RLVR.

\mypara{Evaluation Results}
Our main evaluations and analyses are built on math-reasoning datasets.
Under a white-box setting, our experiments show that DIBA consistently outperforms a strong suite of transferred baselines in our RLVR setting, including calibrated and self-generated variants designed to better match the RLVR generation process, with the clearest gains appearing in the high-precision regime.
Across our main evaluations, DIBA achieves an order of magnitude better TPR@0.1\%FPR and strong ranking performance with AUC around 0.8.
At the same time, we reveal an important boundary: detectability is strongest when RLVR induces observable reward change on a prompt, and remains viable even when the model never solves the task but its behavior shifts; however, it degrades when the base model already achieves consistent success, leaving little room for RLVR to alter behavior.

Under a practical grey-box setting, transfer across model sizes is often robust under the same RLVR algorithm, while cross-algorithm transfer is less stable.
Nevertheless, our results suggest that with carefully chosen shadow data and an RLVR recipe, DIBA can still provide practically useful auditing performance with over 0.7 AUC in white-box settings.
We further extend the analysis to different modalities and verifiable tasks.
For modality transfer, we train vision-language models on a dataset with visual inputs and find that the related signals persist (AUC around 0.7).
For task transfer, we apply DIBA to instruction-following and tool-using datasets, where clear signals persist (AUC around 0.7).
These results provide preliminary evidence for broader applicability beyond math reasoning.
Finally, we study possible mitigations and find that post-hoc output transformations can reduce the externally observable text-level auditing signal, whereas standard regularization offers limited protection in our settings.

In summary, we make the following contributions:
\begin{itemize}[leftmargin=*]
\item We present a systematic study of white-box query-level membership auditing in RLVR, showing that exposure risk in this setting is better associated with prompt-conditioned behavioral traces than with fixed-target fitting.
\item We propose DIBA, a comparative auditing framework that estimates exposure signal from reward-side change and policy-side behavioral change relative to a reference model.
\item We conduct extensive evaluations across multiple settings, including distribution shift, algorithm/model mismatch, and vision-language extensions, and characterize when RLVR auditing succeeds and where it breaks down.
\end{itemize}

\section{Related Work}\label{sec:related}

\mypara{Reinforcement Learning for LLM}
Current RL algorithms can be mainly divided into two parts: preference-based RL and Reinforcement Learning with Verifiable Rewards (RLVR).
Regarding preference-based RL, it is featured by explicitly or implicitly deriving the reward from pairwise data.
Two representative methods are PPO~\cite{bai2022training} and DPO~\cite{rafailov2023direct}.
Proximal Policy Optimization (PPO)~\cite{schulman2017proximal} is a popular actor-critic algorithm used to fine-tune language models with feedback from a reward model. 
To simplify the complexity and save the training cost of PPO, Direct Preference Optimization (DPO)~\cite{rafailov2023direct} was introduced as a more direct method. 

RLVR is characterized by computing the reward by a rule-based verification function and no longer needs the preference data.
It enables the LLMs to have the capability of test-time scaling, thus having longer reasoning paths and stronger reasoning ability~\cite{shao2024deepseekmath,yu2025dapo,liu2025thought,nie2026attnpo,bai2026ttvs}.
Specifically, it first samples a group of outputs for a given query and computes the advantage of each response for updating the policy.
Since the advantages are computed using rule-based reward, e.g., exact match of math answer or execution results of code, RLVR can significantly reduce the required resources.
We will introduce the RL algorithms on LLMs in \Cref{sec:preliminary}.

\mypara{Membership Inference on Fixed-target Supervision}
We parameterized the LLM by a policy ($\pi_\theta$) that generates different actions (responses $o$) given certain states (query $q$).
The base model is referred to as $\pi_\text{ref}$.

In the pre-training stage, adversaries seek to determine whether a given text or paragraph (e.g., a Wikipedia passage) was present in the pre-training corpus~\cite{shi2023detecting,xie2024recall,mattern2023membership,zhang2024min,carlini2021extracting,wang2025recall}. 
Studies in this setting typically adopt LLMs with identifiable data sources, such as GPT-2~\cite{radford2019language} and Pythia~\cite{biderman2023pythia}. 
Several representative methods shown in MIAs at this stage primarily exploit the model’s increased fluency or lower perplexity on data it has seen during pre-training.

In the SFT stage, the goal is to infer whether a specific question-answer pair $(q, o)$, comprising query $q$ and the response $ o$, is included in the SFT dataset. 
Attacks here mainly leverage the model’s tendency to memorize and overfit to seen training samples, leading to distinguishable behavioral patterns on member inputs~\cite{fu2023practical,ran2025lora, dong2024generalization}.
The methods on pre-training data can be seamlessly transferred by replacing the probability term $\pi_\theta(q)$ with the conditional probability $\pi_\theta(o|q)$.
Additionally, we list several methods optimized for SFT data.
\begin{itemize}[leftmargin=*]
    \item \textit{SPV-MIA~\cite{fu2023practical}:}
    It first computes the neighborhood score by $ \text{Score}_{\text{NEI}}(q, \pi_{\theta}) = \frac{1}{N}\sum\log \pi_\theta(\hat{q}) - \log \pi_{\theta}(q)$, where the $\hat{q}$ is the collection of multiple paraphrased version of query $q$ and $N$ is the number of the samples.
    Then, it trains a self-prompted model $\pi_{sp}$ to calibrate the neighborhood score function, i.e.,
    \[\text{Score}_{\text{SPV-MIA}} = \text{Score}_{\text{Nei}}(q,\pi_{sp}) - \text{Score}_{\text{Nei}}(q,\pi_{\theta})\]

    \item \textit{LoRA-Leak~\cite{ran2025lora}:}
    It enhances the MIA score by calibrating with the base model just as previous calibration techniques~\cite{watson2021importance,carlini2022membership} do.
    Compared to using the score derived only from the fine-tuned model, calibrating the score with the base model can make the membership signal more distinguishable.
\end{itemize}
Prior work typically targets fixed supervised objects. 
In prior reference-based MIAs~\cite{watson2021importance,carlini2022membership}, the reference model calibrates the likelihood of a fixed target string under the null hypothesis. 
In DIBA, no such target exists; the reference model provides the counterfactual baseline, and the signal emerges from behavioral divergence between the two policies.

\mypara{Membership Inference on Preference Data}
Regarding the preference-based RL,  the goal is to train a reward model and update the policy to assign a higher probability to preferred responses. 
Adversaries aim to determine whether particular ``query-preferred-rejected''  data $(q, o_\text{prefer}, o_\text{reject})$, comprising query $q$, a preferred response $ o_\text{prefer}$, and a rejected response $o_\text{reject}$, is used during reward modeling or policy optimization. 
The corresponding MIA targets triplets and leverages the gap in likelihoods between chosen and unchosen responses to infer membership. 
A representative approach is PREMIA~\cite{feng2024exposing}, which quantifies the model's advantage for the preferred response over the rejected one. 
\[\text{Score}_{\text{PREMIA}} = \frac{\pi_\theta(o_\text{prefer} |q)}{\pi_\text{ref}(o_\text{prefer} |q)} - \frac{\pi_\theta(o_\text{reject} |q)}{\pi_\text{ref}(o_\text{reject} |q)}\]
MIAs in this context exploit the model’s stronger preference for the chosen response over the rejected one, often reflected in probability gaps or response rankings.

\mypara{Membership Inference on Generative Models}
Compared to existing MIAs on LLM, the challenge in auditing RLVR is more relevant to MIAs on generative models, where the attacker cannot directly rely on class-confidence signals or fixed supervised targets. 
LOGAN~\cite{hayes2017logan} shows that, for GAN-based generative models, membership can be inferred from distributional discrepancies captured by the generation process rather than from standard classification scores. 
More recently, SecMI~\cite{duan2023diffusion} demonstrates that diffusion models are also vulnerable to membership inference, using step-wise estimation errors across the denoising process as the attack signal. 
Our work is closest in spirit to membership auditing for generative models, but differs in both target and observable. 
\section{Preliminaries and Auditing Setup}
\label{sec:preliminary}

In this section, we introduce the notation and formulations of representative RLVR algorithms used in foundation-model post-training, including GRPO and DAPO. 
We also clarify the auditing objects considered in this work, since our setting differs from fixed-target membership inference in pre-training, SFT, and preference-based alignment.

\mypara{Notation}
We denote the reference policy (e.g., the initial model) by $\pi_{\text{ref}}$, the current policy by $\pi_{\theta}$, and the sampling policy by $\pi_{\theta_{\text{old}}}$. 
The prompt for sampling is represented by $q$, and the sampled response is denoted by $o$.

\mypara{GRPO~\cite{shao2024deepseekmath}}
GRPO is a recent RL algorithm designed to retain the benefits of online reinforcement learning while significantly reducing the resource burden of PPO. 
Instead of learning a value function for baseline estimation, GRPO uses a group-based baseline. 
Specifically, for each prompt $q$, it samples multiple responses $\{o_1, \dots, o_G\}$ from the current policy, scores them with a rule-based function (e.g., whether the response correctly solves a math problem), and normalizes the rewards by subtracting the group mean and dividing by the group standard deviation. 
Formally, the step-level sparse advantage of each token in the response is computed as
$
\hat{A}_{i,t}=\frac{r_i-\mathrm{mean}(\mathbf{r})}{\mathrm{std}(\mathbf{r})},
$
where $r_i$ is the reward of response $o_i$.
The GRPO objective is:
\begin{equation}\label{eq:lossgrpo}
\begin{split}
\mathcal{J}(\theta) = &\mathbb{E}_{[q, \{o_i\}^G\sim \pi_{\theta_{\text{old}}}]}  \frac{1}{G} \sum_{i=1}^G \frac{1}{|o_i|}\sum_{t=1}^{|o_i|} \\\Big( &\frac{\pi_\theta(o_{i,t} \mid q, o_{i,<t})}{\pi_{\theta_{\text{old}}}(o_{i,t} \mid q, o_{i,<t})} \hat{A}_{i,t} 
- \beta D_{\text{KL}}\left( \pi_{\text{ref}}, \pi_\theta \right) \Big) ,
\end{split}
\end{equation}
where $D_{\text{KL}}$ is the KL divergence and $\beta$ controls the strength of regularization.
For simplicity, we do not explicitly show the standard PPO-style clipping term in the first term.

\mypara{DAPO~\cite{yu2025dapo}}
DAPO is a widely adopted variant of GRPO that improves training stability through several modifications. 
It introduces clip-higher, which decouples the upper and lower clipping bounds to stabilize policy updates. 
The rebalancing act mechanism gives longer sequences greater influence in the gradient computation. 
Dynamic sampling filters out prompts with perfect or zero accuracy within each batch, ensuring non-zero advantages and meaningful learning signals. 
In addition, DAPO uses overlong reward shaping to penalize excessively long responses for efficiency, and removes the explicit KL-divergence regularization term. 
The full optimization objective is shown in \Cref{app:preliminary}.

\mypara{RLVR and Auditing Implications}
RLVR differs from SFT and preference-based RL in that it reuses prompts while generating trajectories on-policy and scoring them with verifiable rewards. 
This difference has two implications for auditing. 
First, unlike SFT, RLVR does not repeatedly optimize toward a single reference string, so fitting of a fixed answer is no longer the most natural observable. 
Second, unlike preference-based RL such as PPO or DPO with pairwise preference data, RLVR derives learning signals from non-pairwise verifiable outcomes, such as correctness checks or execution results. 
These properties motivate query-level auditing based on post-training behavioral traces, rather than fixed-target scoring.

\section{Threat Model}
\label{sec:threat}

We consider the commonly used white-box post-training scenario studied in prior work, where the victim starts from a publicly available pretrained model, performs RLVR fine-tuning on a private prompt set, and releases the resulting model weights ($\pi_\theta$). 
Our goal is to audit whether candidate prompts have participated in RLVR optimization and whether such participation leaves an observable post-training trace in the released model.

\mypara{Auditor's Goal}
The auditor ultimately seeks to infer whether a candidate prompt participated in RLVR training.
This corresponds to compliance auditing or contamination detection.
However, in RLVR, prompts that are already well handled by the base model often yield little remaining learning signal and thus induce only weak additional updates~\cite{yu2025dapo,nan2025ngrpo}.
Accordingly, although training participation is the underlying object of interest, the prompts of greatest practical relevance for auditing are those that undergo non-trivial policy improvement, since these are more likely to leave security-relevant traces in the released model.

\mypara{Auditor's Knowledge and Capability}
We consider an inference-only auditor in a white-box setting. 
The auditor can access the released model and query token-wise logits, but does not use gradients, optimizer states, or the victim's training logs. 
We assume the auditor knows that an RLVR algorithm was used, but not the exact hyperparameters or recipe.

Our auditing compares the released model against a reference checkpoint. 
We assume the auditor has access to the exact base model or a near-equivalent reference model. 
This assumption is reasonable in white-box ecosystems for two reasons: derivative release typically requires disclosing the base family~\cite{llama-license}, and recent work on model lineage and fingerprinting~\cite{shang2026attesting,pasquini2025llmmap} suggests that the foundation can often be identified even when it is not explicitly advertised.

We also assume that the auditor can verify whether a sampled response satisfies the RLVR objective. 
This assumption is appropriate for objective RLVR tasks with verifiable outcomes, such as mathematical reasoning or code execution, where correctness can be checked by a generic verifier rather than proprietary hidden labels.
The task does not directly extend to subjective preference-based alignment.

\mypara{Threat Model A: Auditing Upper Bound}
In the upper-bound setting, the auditor additionally possesses labeled shadow samples drawn from the same task distribution as the victim's RLVR data, while remaining disjoint from the actual training set. 
This setting measures the intrinsic detectability of RLVR exposure under strong observability.
Our main experiments instantiate Threat Model A. 

\mypara{Threat Model B: Practical Grey-box Audit}
In the practical grey-box setting, the auditor does not have the same-distribution labeled samples. 
Instead, they use public shadow datasets with similar task types, and must tolerate mismatch in data source, model initialization, or RLVR algorithm. 
This setting measures how much auditing signal survives under realistic approximation and transfer.
Our shadow-data and transfer experiments instantiate Threat Model B.
\section{Methodology}
We propose DIBA, a query-level exposure auditor for RLVR. 
Rather than scoring fitting of a fixed target response, DIBA estimates whether RLVR leaves an observable optimization trace on a prompt relative to a reference checkpoint.

\subsection{Motivation: Observable Optimization Traces in RLVR}
\mypara{SFT Gradient}
In SFT, each training example is associated with a fixed target response. 
Concretely, the SFT gradient can be written as:
\begin{equation}
\begin{split}
     \nabla_{\theta}\mathcal{J}(\theta)   = \mathbb{E}_{[q , o \sim \mathcal{D}]}  \sum_{t=1}^{|o|} 
     \nabla_{\theta}\log \pi_\theta(o_{t} | q, o_{<t}),
\end{split}
\end{equation}
where $\mathcal{D}$ is the fine-tuning dataset.  
This objective repeatedly pushes probability mass toward the specific response $o$ paired with prompt $q$.
As a result, response-level signals are a basis for membership auditing.

\mypara{RLVR Gradient}
RLVR differs in that responses are generated on-policy and rewarded according to a verifiable outcome, rather than matched to a fixed reference string. 
In GRPO-style RLVR, the policy gradient takes the form
\begin{equation}
\begin{split}
     \nabla_{\theta}\mathcal{J}(\theta)   = \mathbb{E}_{[q , \{o_i\}^G \sim \pi_{\theta_{old}}]} 
     \frac{1}{G}&\sum_{i=1}^G\frac{1}{|o_i|} \sum_{t=1}^{|o_i|} \\m_{i,t} 
     &\nabla_{\theta}\log \pi_\theta(o_{i,t} | q, o_{i,<t}),
\end{split}
\end{equation}
where the gradient coefficient $m_{i,t}$ is defined as follows:
\begin{equation}\label{eq:coef}
    m_{i,t} =  \underbrace{\hat{A}_{i,t}}_{\text{Advantage Term}}+ \beta \underbrace{(\frac{\pi_{ref}(o_{i,t}|q,o_{i,<t})}{\pi_{\theta}(o_{i,t}|q,o_{i,<t})} - 1)}_{\text{Divergence Term}}.
\end{equation}
Because RLVR does not repeatedly optimize toward a single target response for each prompt, fixed-target response scoring is less directly aligned with the training mechanism. 
Instead, the more direct observables are how RLVR changes the model's expected reward on a prompt and how far the fine-tuned policy departs from the pre-RLVR reference policy.
For DAPO, which removes the explicit KL regularization term, divergence is better viewed as an empirical policy-shift feature rather than a direct instantiation of the same regularizer.

\mypara{New Leakage Channel}
This gradient view motivates two prompt-level signals for RLVR auditing.
Specifically, the coefficient \(m_{i,t}\) decomposes RLVR updates into two drivers:
\begin{itemize}[leftmargin=*]
    \item The advantage term \(\hat{A}_{i,t}\) amplifies trajectories that achieve higher verifiable reward. 
    For prompts that are not already saturated under the base model, repeated RLVR optimization can lead to measurable reward improvement.
    \item The divergence term constrains updates toward the reference policy \(\pi_{\mathrm{ref}}\), yet empirical learning under RLVR is often accompanied by deviations from \(\pi_{\mathrm{ref}}\) on trajectories. 
    Thus, RLVR participation may leave measurable token-level probability shifts between \(\pi_{\mathrm{ft}}\) and \(\pi_{\mathrm{ref}}\).
\end{itemize}
This gradient view suggests two observable traces of RLVR exposure and motivates our use of reward improvement and policy divergence as observable proxies for prompt-level membership auditing.
\begin{table}[]
\centering
\caption{Comparison with baselines. 
To avoid a strawman comparison, we consider three baseline settings, including static-target, calibrated, and self-generated settings.
We use TPR to represent TPR@0.1\%FPR.}
\label{tab:baselines_main}
\resizebox{\linewidth}{!}{\begin{tabular}{c|cc|cc|cc|cc}
\midrule
\multirow{2}{*}{Method} & \multicolumn{2}{c|}{GRPO-3B} & \multicolumn{2}{c|}{DAPO-3B} & \multicolumn{2}{c|}{GRPO-7B} & \multicolumn{2}{c}{DAPO-7B} \\ \cmidrule{2-9} 
 & AUC $\uparrow$ & TPR $\uparrow$& AUC $\uparrow$& TPR $\uparrow$& AUC $\uparrow$& TPR $\uparrow$& AUC $\uparrow$& TPR $\uparrow$\\ \midrule
LOSS~\cite{carlini2021extracting} & 0.568 & 0.006 & \textbf{0.562} & \textbf{0.008} & \textbf{0.575} & 0.002 & \textbf{0.582} & 0.000 \\
$\hookrightarrow$ \textit{Calibrate} & 0.513 & \textbf{0.009} & 0.527 & 0.007 & 0.553 & \textbf{0.004} & 0.571 & 0.005 \\
$\hookrightarrow$ \textit{Self-gen} & \textbf{0.571} & 0.002 & 0.504 & 0.007 & 0.554 & 0.003 & 0.546 & \textbf{0.011} \\ \midrule
Zlib~\cite{carlini2021extracting} & 0.536 & 0.002 & 0.530 & 0.003 & 0.541 & 0.000 & 0.543 & 0.001 \\
$\hookrightarrow$ \textit{Calibrate} & 0.519 & \textbf{0.004} & 0.525 & \textbf{0.010} & 0.527 & 0.001 & 0.541 & 0.003 \\
$\hookrightarrow$ \textit{Self-gen} & \textbf{0.588} & 0.002 & \textbf{0.588} & 0.003 & \textbf{0.571} & \textbf{0.003} & \textbf{0.583} & \textbf{0.016} \\ \midrule
Min-K\%~\cite{zhang2024min} & 0.524 & 0.000 & 0.548 & 0.003 & 0.519 & 0.002 & 0.521 & 0.002 \\
$\hookrightarrow$ \textit{Calibrate} & \textbf{0.547} & \textbf{0.016} & \textbf{0.567} & 0.006 & 0.524 & \textbf{0.008} & 0.517 & 0.003 \\
$\hookrightarrow$ \textit{Self-gen} & 0.525 & 0.000 & 0.522 & \textbf{0.005} & \textbf{0.525} & 0.002 & \textbf{0.526} & \textbf{0.004} \\ \midrule
Entropy & \textbf{0.586} & 0.005 & 0.586 & 0.005 & 0.588 & 0.005 & 0.584 & 0.002 \\
$\hookrightarrow$ \textit{Calibrate} & 0.533 & 0.006 & 0.512 & 0.004 & \textbf{0.598} & \textbf{0.008} & 0.579 & 0.006 \\
$\hookrightarrow$ \textit{Self-gen} & 0.553 & \textbf{0.009} & \textbf{0.606} & \textbf{0.011} & 0.533 & 0.007 & \textbf{0.645} & \textbf{0.008} \\ \midrule
Neighbor~\cite{mattern2023membership} & 0.523 & 0.005 & 0.518 & 0.004 & 0.514 & 0.002 & 0.518 & 0.006 \\ \midrule \midrule
\textbf{Ours (DIBA)} & \textbf{\color[HTML]{1F77B4}0.713} & \textbf{\color[HTML]{1F77B4}0.073} & \textbf{\color[HTML]{1F77B4}0.835} & \textbf{\color[HTML]{1F77B4}0.154} & \textbf{\color[HTML]{1F77B4}0.749} & \textbf{\color[HTML]{1F77B4}0.084} & \textbf{\color[HTML]{1F77B4}0.819} & \textbf{\color[HTML]{1F77B4}0.107} \\ \midrule
\end{tabular}
}\end{table}
\subsection{Transfer of Likelihood-based MIAs to RLVR}
To evaluate how well existing fixed-target MIAs transfer, we benchmark representative MIAs for LLMs and to test whether their standard instantiations remain effective under RLVR settings.

\mypara{Experimental Setup}
We use the same trained models as in \Cref{sec:eval}; implementation details are deferred to \Cref{app:base-imple}. 
To avoid a strawman comparison, we consider three baseline settings, thus deriving 13 baseline variants:
\begin{itemize}[leftmargin=*]
    \item \textit{Static-target Setting.} For each prompt \(q\), we assume the adversary has access to the SFT reference answer \(o^\star\), and computes scores on the fixed pair \((q, o^\star)\) from existing MIA methods.
    This corresponds to the conventional SFT-style membership setting.
    \item \textit{Calibrated Setting.} We further consider a calibration-based variant~\cite{ran2025lora}.
    For a sampled target string \(o\), the calibrated score takes the form
\(
s_{\mathrm{cal}}(q,o)=s_{\pi_{ft}}(q,o)-s_{\pi_{ref}}(q,o),
\)
where \(s\) denotes the corresponding likelihood-based metric.
    This tests whether reference-based calibration improves transfer to RLVR.
    \item \textit{Self-generated Setting.} For each prompt \(q\), we sample \(N\) responses \(\{o_i^{ft}\}_{i=1}^N\) from the fine-tuned model \(\pi_{ft}(\cdot \mid q)\).
    For each sampled response, we compute the corresponding MIA score under \(\pi_{ft}\), calibrated by the reference model \(\pi_{ref}\), and then aggregate these sample-level scores into a prompt-level score by averaging across the \(N=8\) rollouts. 
    This gives the baselines access to adaptive trajectories that better match the RLVR generation process.
\end{itemize}

\mypara{Evaluation Results}
These baselines are intentionally strong transfers, but they remain misaligned with the RLVR auditing object because they score fixed strings.
As shown in \Cref{tab:baselines_main}, these baselines transfer substantially worse than DIBA in our RLVR setting, with limited separability overall and especially weak performance in the low-FPR regime. 
This trend persists even under the self-generated setting, where the baselines are given adaptive trajectories sampled from the fine-tuned model itself. 
These results suggest that, although existing MIAs remain a meaningful comparison point, fixed-target response scoring is not the most informative observable for prompt-level auditing in RLVR. 
This motivates moving from response-level likelihood signals to query-level estimates of reward improvement and policy divergence.

\subsection{Divergence-in-Behavior Auditing}
We propose \textbf{D}ivergence-\textbf{i}n-\textbf{B}ehavior \textbf{A}uditing (DIBA), a query-level auditing framework tailored to RLVR, where the auditing target is a prompt \(q\) without any provided response. 
The core idea is to estimate how RLVR fine-tuning changes model behavior relative to the base policy along the two observable axes suggested by the RLVR gradient view: reward-side improvement and policy-side divergence.

\begin{figure}[t]
    \centering
    \includegraphics[width=1\linewidth]{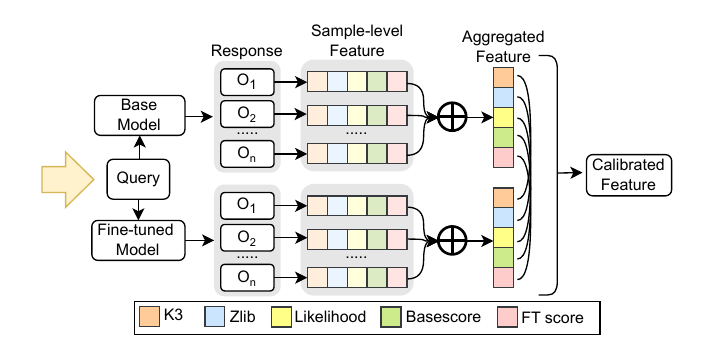}
    \caption{Overview of DIBA's feature construction.}
    \label{fig:diba-method}
\end{figure}

\mypara{Preparation}
Let \(\pi_{\mathrm{ft}}\) denote the RLVR fine-tuned policy and \(\pi_{\mathrm{ref}}\) the corresponding base policy. 
For each query \(q\), we draw \(N\) stochastic rollouts from each model:
\[
{o^{\mathrm{ft}}_i}_{i=1}^N \sim \pi_{\mathrm{ft}}(\cdot\mid q),\quad
{o^{\mathrm{ref}}_i}_{i=1}^N \sim \pi_{\mathrm{ref}}(\cdot\mid q).
\]
These samples allow us to estimate reward-level performance and token-level distributional shifts~\Cref{fig:diba-method}. 
In practice, a sampling size of 4--8 is sufficient to provide stable prompt-level estimates.

\mypara{Feature - Reward-side Signal}
Let \(\mathcal{R}\) be the rule-based evaluator, e.g., mathematical verification. 
We quantify how RLVR changes the model's expected reward on \(q\), which serves as a prompt-level proxy for the advantage-driven component of RLVR updates. 
Specifically, we compute the fine-tuned score and base score as Monte Carlo estimates:
\[
s^{\text{Adv}}_{q} =
[
\underbrace{\frac{1}{N}\sum^N_{i=1}\mathcal{R}(o^{\text{ft}}_i)}_{\text{FT Score}},
\underbrace{\frac{1}{N}\sum^N_{i=1}\mathcal{R}(o^{\text{ref}}_i)}_{\text{Base Score}}
].
\]
Intuitively, prompts that induce nontrivial optimization during RLVR are more likely to exhibit measurable reward-side change relative to the base model. 
This signal primarily contributes to overall separability, but, as we analyze later, can be vulnerable to false positives on inherently easy prompts where both base score and FT score are high.

\begin{figure*}[t]
    \centering
    \includegraphics[width=0.9\textwidth]{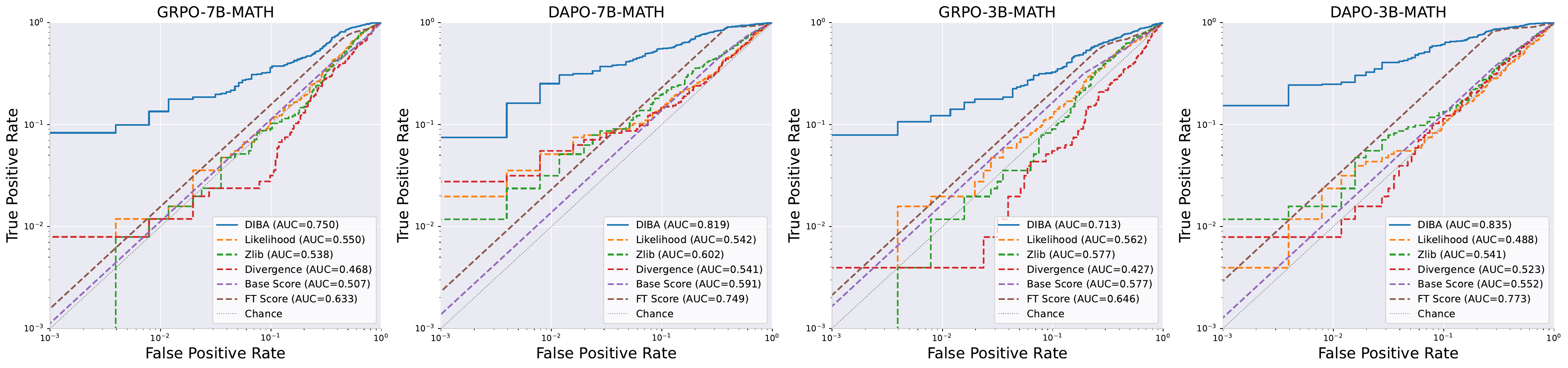}
    \caption{AUC and log-scale ROC curves for DIBA on different models and training algorithms. The ROC and AUC of the component features are calibrated on the train set.
    All results are obtained via 5 runs on different random seeds.}
    \label{fig:roc-main}
\end{figure*}

\mypara{Feature - Policy-side Signal}
Reward-side change alone is insufficient for high-precision auditing. 
RLVR can also induce token-level probability shifts relative to \(\pi_{\mathrm{ref}}\); these shifts reflect policy-side behavioral drift and provide complementary evidence in the low-FPR regime.

\begin{itemize}[leftmargin=*]
    \item \textit{K3 Divergence}: We measure how far \(\pi_{\mathrm{ft}}\) drifts from \(\pi_{\mathrm{ref}}\) on its own rollouts. 
    For each sampled response \(o^{\mathrm{ft}}_i\), we compute a token-level divergence score using the unbiased \texttt{k3}~\cite{kl_approx} estimator of KL divergence:
\[ s^{\text{KL}}_{q,o_i^{\text{ft}}} =\frac{1}{|o_i|}\sum_{t=1}^{|o_i|}\frac{\pi_{{ref}}(o_{i,t} \mid q, o_{<t})}{\pi_{ft}(o_{i,t} \mid q, o_{<t})} - \log \frac{\pi_{{ref}}(o_{i,t} \mid q, o_{<t})}{\pi_{ft}(o_{i,t} \mid q, o_{<t})} - 1, \]    and then aggregate across rollouts:
    \[
    s^{\mathrm{Div}}(q)=\frac{1}{N}\sum_{i=1}^{N} s^{\mathrm{KL}}_{q,o^{\mathrm{ft}}_i}.
    \]
    This feature captures the magnitude of policy shift and is referred to as \texttt{Divergence} in the following text.

    \item \textit{Confidence Shift}: In addition to direct divergence estimation, we include an empirical measure of confidence shift inspired by likelihood-ratio calibration~\cite{carlini2022membership}. 
    For rollouts from \(\pi_{\mathrm{ft}}\), we compute
    \[
    s^{\mathrm{Likelihood}}_q
    =
    \frac{1}{N}\sum_{i=1}^N
    \left[
    \log \pi_{\mathrm{ft}}(o_i^{\mathrm{ft}} \mid q)
    -
    \log \pi_{\mathrm{ref}}(o_i^{\mathrm{ft}} \mid q)
    \right],
    \]
    and a redundancy-robust variant normalized by compressed length~\cite{carlini2021extracting}:
    \[
    s^{\mathrm{Zlib}}_q
    =
    \frac{1}{N}\sum_{i=1}^N
    \frac{
    \log \pi_{\mathrm{ft}}(o_i^{\mathrm{ft}} \mid q)
    -
    \log \pi_{\mathrm{ref}}(o_i^{\mathrm{ft}} \mid q)
    }{
    |\mathrm{zlib.compress}(o_i^{\mathrm{ft}})|
    }.
    \]

    Importantly, our use of \(\pi_{\mathrm{ref}}\) is not merely calibration on a fixed target string. 
    Since RLVR lacks ground-truth targets, DIBA relies on comparative sampling to estimate prompt-conditioned behavioral change. 
    In this sense, \(\pi_{\mathrm{ref}}\) serves as an active baseline policy rather than a static filter.
\end{itemize}

\mypara{Prediction}
For each prompt \(q\), we concatenate the above prompt-level features \((s^{\text{Adv}}_{q}, s^{\mathrm{Div}}_q, s^{\mathrm{Likelihood}}_q, s^{\mathrm{Zlib}}_q)\) into a feature vector. 
We then train a simple stacking predictor, which ensembles random forest and logistic regression with standardization, to output a prompt-level auditing score. 
When binary decisions are needed, this score is thresholded to predict whether the prompt is likely exposed under RLVR.

We evaluate DIBA from three perspectives: in-distribution detectability under an upper-bound auditing setting, transferability under practical grey-box mismatch, and empirical boundaries such as already-mastered prompts, algorithm/model mismatch, and multimodal extension.

\section{Attack Evaluation}
\label{sec:eval}

\mypara{Models}
For RLVR, we select two widely used models: {Qwen-2.5-7B-Instruct} and {Qwen-2.5-3B-Instruct}~\cite{yang2024qwen2}. 
For the multi-modal tasks, we utilize Qwen-2.5-7B-Instruct-VL~\cite{yang2024qwen2}.
We focus exclusively on the Qwen series because recent work~\cite{gandhi2025cognitive} demonstrates that Qwen family exhibits significantly stronger emergent reasoning capabilities under RLVR compared to the Llama family, whose pre-trained policies often lack the foundational cognitive behaviors that is necessary for effective reinforcement learning. 
As a result, Qwen has become the de facto standard in post-training research~\cite{yu2025dapo, cui2025entropy, shao2025spurious}, and nearly all recent advances in RLVR are evaluated or developed using Qwen variants.

\mypara{Datasets}
Regarding the dataset, we use different parts of the MATH dataset~\cite{hendrycksmath2021} and its variant~\cite{numina_math_datasets}, the most widely adopted benchmark in RLVR research, which comprises competition-level math problems, to construct our membership auditing evaluation. 
Specifically, to minimize the risk of potential data leakage from previous training, we draw both member and non-member prompts exclusively from the test split of the original dataset. 

Beyond our main evaluation and analysis on math reasoning, we extend DIBA to a visually-grounded math dataset~\cite{lu2021inter}, an instruction-following dataset~\cite{ye2025multi}, and a tool-using dataset~\cite{liu2024apigen} to test whether the auditing signal generalizes across domains.
  
\mypara{Training Setups}
To conduct the RL training, we utilize verl~\cite{sheng2024hybridflow} as the training framework, which greatly accelerates the rollout and training stages.
Specifically, we consider 2 widely used RLVR algorithms, i.e., GRPO~\cite{shao2024deepseekmath} and DAPO~\cite{yu2025dapo}.
Regarding the training hyperparameters, we use the recommended parameters in verl library and show the details in \Cref{app:hyper}.
The model is trained on 4 NVIDIA H100 GPUs, where a full 60-epoch run takes around 18 hours to complete. 

\mypara{Evaluation Metrics}
We report multiple metrics to comprehensively assess membership auditing performance following previous work~\cite{carlini2022membership,fu2023practical}: balanced accuracy, AUC, log-scaled ROC curves, and TPR@0.1\%FPR. 
All quantitative results are obtained via 5 runs on different random seeds.
The predictor is trained and evaluated on distinct splits of member and non-member prompts. 
Numerical metrics (AUC, balanced accuracy, TPR@FPR) are reported as the mean across five independent runs. 
The \textit{implementation details} are shown in \Cref{app:hyper}.

\subsection{Main Evaluation}

\Cref{fig:roc-main} presents the attack performance of DIBA and its component features, several of which correspond to enhanced baseline signals inspired by LoRA-Leak~\cite{ran2025lora}. 
Overall, DIBA outperforms all individual baselines and feature-based variants in both AUC and TPR@0.1\%FPR, demonstrating the effectiveness of our integrated approach.  
Across different RLVR algorithms, models fine-tuned with DAPO exhibit higher vulnerability to membership auditing compared to those trained with GRPO. 
This increased exposure may stem from DAPO’s more effective training dynamics and stronger learning capacity, leading to larger behavioral shifts from the base model, a phenomenon we analyze further in \Cref{sec:fit}.
When comparing model sizes, we observe no consistent trend in whether larger models leak more membership information.
This suggests that scale alone does not linearly correlate with privacy risk in the RLVR setting.

Additionally, we observe that the TPR@0.1\%FPR achieved by DIBA is significantly higher by an order of magnitude at least than that obtained by any individual feature alone. 
This substantial gain suggests a synergistic interaction among features, where complementary signals enhance detection power in the high-precision regime. 
We further analyze the contribution and interplay of these features in \Cref{sec:importance}.

\subsection{Auditability Regimes in RLVR}\label{sec:regimes}

RLVR does not leave equally strong auditing signals on all training prompts.
When the KL coefficient $\beta$ is small, policy updates are driven mainly by the advantage term in \Cref{eq:coef}.
In particular, the group-normalized advantage
$\hat{A}_{i,t} = \frac{r_i - \text{mean}(\mathbf{r})}{\text{std}(\mathbf{r})}$
collapses to zero when all sampled responses receive the same reward.
Thus, prompts with saturated outcomes contribute little task-driven update and may leave only weak additional traces in the released model.

Motivated by this observation, we partition prompts into three regimes according to how much behavioral change is observable before and after RLVR:
\begin{itemize}[leftmargin=*]
    \item \textit{Already-mastered:} prompts on which both the base and fine-tuned models achieve an average score of 1.
    \item \textit{Never-learned:} prompts on which both models achieve an average score of 0.
    \item \textit{Actively-learned:} the remaining prompts, on which RLVR is associated with non-trivial behavioral change.
\end{itemize}

For each regime, we train a separate DIBA predictor.
As shown in \Cref{tab:hard-main}, the \textit{Actively-learned} regime is consistently the easiest to audit, substantially outperforming the other two regimes in both AUC and TPR@0.1\%FPR.
By contrast, the \textit{Already-mastered} and \textit{Never-learned} regimes are markedly harder.
This suggests that RLVR membership is most auditable when training continues to produce prompt-specific learning traces, and becomes much harder to detect when a prompt is either already solved or never effectively improved.

Interestingly, the \textit{Already-mastered} regime is the hardest.
A plausible explanation is that these prompts exhibit minimal visible change across checkpoints.
Using Rouge-L and 3-gram overlap (\Cref{fig:hard_sim}), we find that responses in the \textit{Already-mastered} regime remain substantially more similar before and after RLVR than those in the other regimes.
This stability leaves only weak incremental evidence even for true training prompts, thereby reducing auditability.

Overall, these results identify an important boundary of RLVR membership auditing:
DIBA is strongest when RLVR leaves non-trivial prompt-specific learning traces, and weakest when training induces little additional behavioral change beyond the reference model.

\begin{table}[t]
\centering
\caption{Auditing performance by prompt regime. The reported AUC and TPR@0.1\%FPR are averaged over 5 runs.}
\label{tab:hard-main}
\large
\resizebox{\linewidth}{!}{
\begin{tabular}{c|c|ccc|ccc}\toprule
\multirow{2}{*}{Model} & \multirow{2}{*}{Split} & \multicolumn{3}{c|}{GRPO} & \multicolumn{3}{c}{DAPO} \\ \cmidrule{3-8}
 &  & \# Samples & AUC & TPR@0.1\%FPR & \# Samples & AUC & TPR@0.1\%FPR \\ \midrule
\multirow{4}{*}{Math-3b} & Never-learned & 108 & 0.687 & 0.047 & 95 & 0.559 & 0.000 \\
 & Already-mastered & 119 & 0.551 & 0.028 & 118 & 0.518 & 0.014 \\
 & Actively-learned & 281 & 0.765 & 0.130 & 294 & 0.912 & 0.282 \\ \cmidrule{2-8}
 & Unified & 508 & 0.713 & 0.073 & 508 & 0.834 & 0.154 \\ \midrule
\multirow{4}{*}{Math-7b} & Never-learned & 76 & 0.661 & 0.064 & 69 & 0.528 & 0.030 \\
 & Already-mastered & 191 & 0.525 & 0.005 & 189 & 0.485 & 0.016 \\
 & Actively-learned & 240 & 0.877 & 0.162 & 250 & 0.937 & 0.280 \\ \cmidrule{2-8}
 & Unified & 508 & 0.750 & 0.084 & 508 & 0.819 & 0.107 \\\bottomrule
\end{tabular}}
\end{table}

\subsection{Effect of Model Fitting}
\label{sec:fit}
We examine how the auditing signal evolves during RLVR training.
We define \textit{train reward} as the accuracy on responses sampled from prompts used in RLVR fine-tuning, reflecting performance on prompts directly exposed to optimization.
In contrast, \textit{test reward} measures accuracy on held-out prompts and serves as a proxy for generalization to unseen inputs.

\begin{figure}[t]
    \centering
    \includegraphics[width=0.9\linewidth]{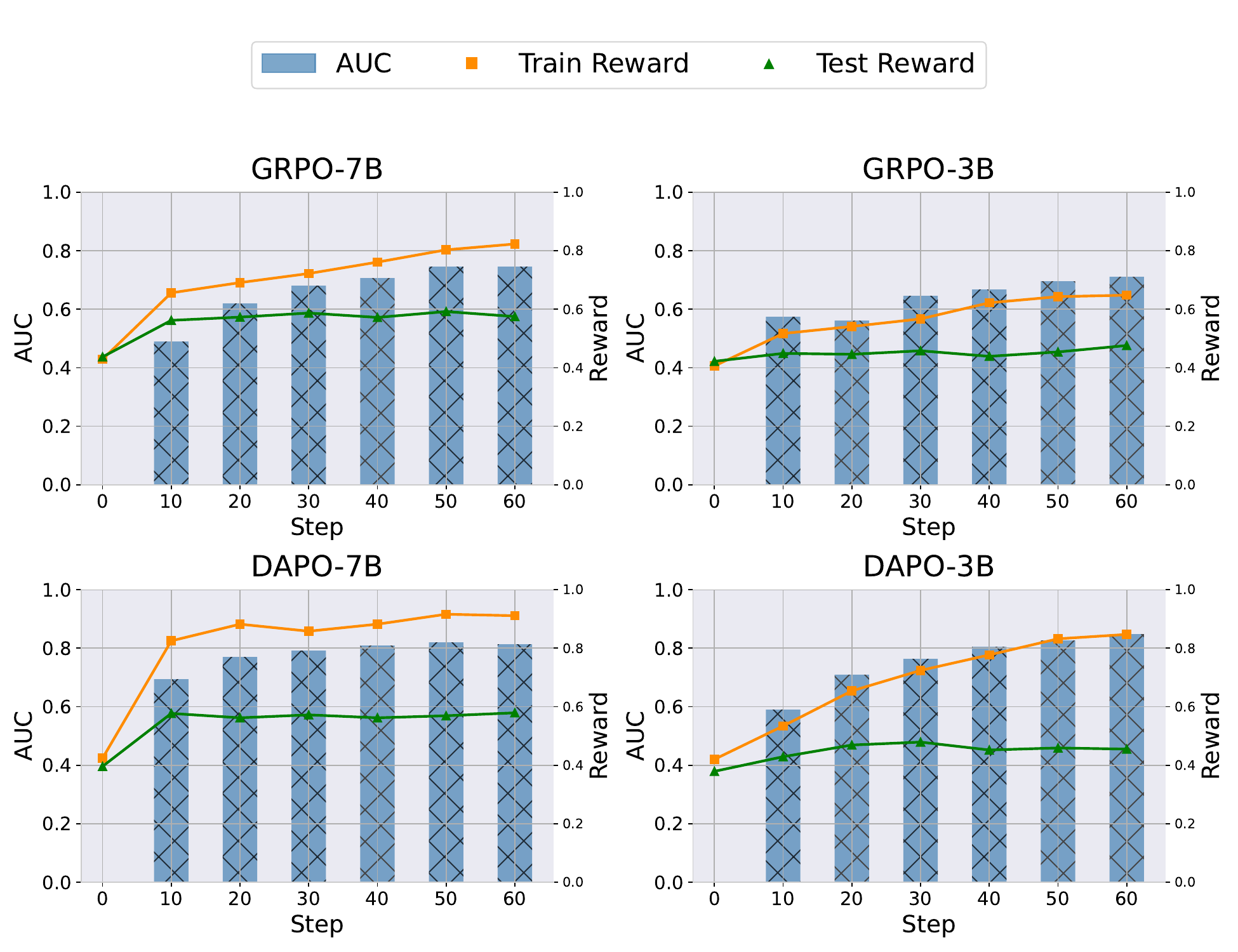}
\caption{Evolution of train reward, test reward, and DIBA AUC during RLVR training. Auditability increases as the train-test reward gap widens.}
\label{fig:ablate_step}
\end{figure}

A key takeaway from \Cref{fig:ablate_step} is that DIBA's auditing signal is not static, but emerges progressively during RLVR training.
Early in training, when the model behaves similarly on training and held-out prompts, the two groups remain difficult to distinguish and DIBA performs near chance.
As RLVR continues, training prompts accumulate stronger prompt-specific behavioral changes, which makes them increasingly separable from held-out prompts.
This observation suggests that DIBA is most sensitive to prompts on which RLVR is still producing nontrivial optimization effects, rather than to prompts whose behavior was already similar before RLVR.

This perspective also helps explain the saturation behavior discussed in \Cref{sec:hardness}.
Once a prompt is handled reliably, further prompt-specific change may diminish because additional reward improvement becomes limited.
In that regime, even true training prompts may leave only weak incremental traces relative to the base model.

Overall, DIBA is most effective when RLVR continues to create observable separation between training and held-out prompts, and less effective when such separation has already saturated or never becomes substantial.

\subsection{Analysis on Feature Importance}\label{sec:importance}
To understand the factors driving auditing in RLVR, we conduct a progressive analysis of DIBA’s feature components, examining their roles across different detection regimes, i.e., from overall ranking to high-precision identification and hard-case behavior.

\mypara{Overall Discriminative Power}
We begin by analyzing DIBA’s performance in terms of AUC, which captures the attack’s overall ability to rank member prompts above non-members without strict false positive constraints.
As shown in \Cref{fig:roc-main}, among all feature components, the \texttt{FT Score}, which measures the improvement in rule-based correctness from base to fine-tuned model, achieves significantly higher AUC than other signals. 
This reflects the central role of reward-driven behavioral change in membership leakage: prompts that induce consistent performance gains during RLVR training are globally more distinguishable. 

However, high AUC alone does not imply practical utility under privacy-sensitive conditions.
For instance, at FPR of 0.1\%, it achieves almost zero TPR in the GRPO-7B-MATH setting, while DIBA achieves a TPR of 0.084, over an order of magnitude higher. 
This mismatch reveals a critical limitation: while FT Score effectively separates populations on average, it lacks the calibration needed to identify the most elusive members under strict thresholds.
\begin{figure}
    \centering
    \includegraphics[width=0.9\linewidth]{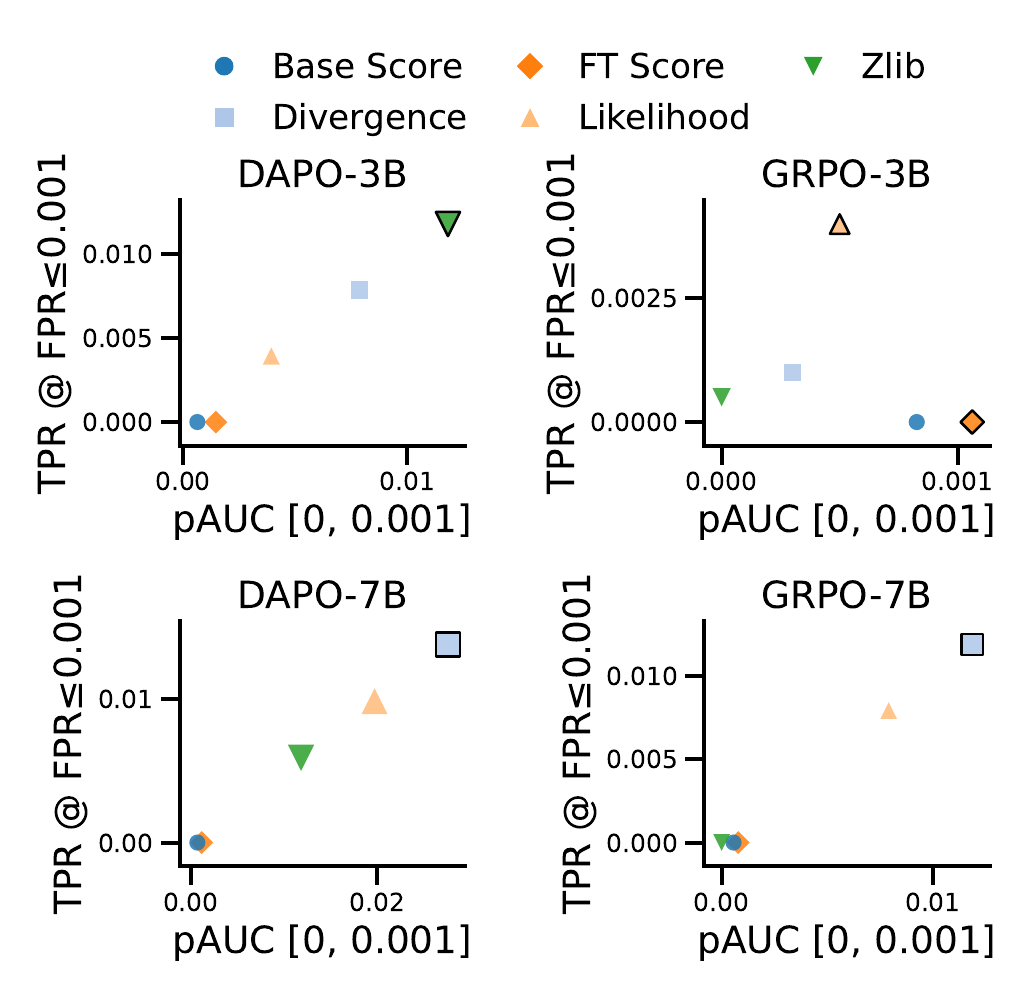}
    \caption{Pareto Front Visualization of Feature Performance: pAUC and TPR@FPR$\leq$0.1\% are on the axes. Marker size reflects the KS statistic (larger = higher), and black rings indicate Pareto-optimal features.}
    \label{fig:Pareto}
\end{figure}

\mypara{Low-FPR Performance}
In privacy-sensitive scenarios, attackers must operate at extremely low false positive rates. 
To evaluate this regime, we adopt a Pareto analysis (\Cref{fig:Pareto}) using pAUC (0–0.1\% FPR) and TPR@threshold, with point size reflecting tail separation (KStail). 
Here, \texttt{Divergence}, which quantifies the KL shift from base to fine-tuned policy, emerges as the most effective single feature, consistently achieving the highest TPR and occupying the Pareto frontier across model variants. 
\texttt{Likelihood} and \texttt{Zlib} provide meaningful but secondary contributions. 
In contrast, \texttt{FT Score} yields zero TPR despite high pAUC, indicating it ranks positives well but fails to generate extreme outlier scores necessary for thresholding. 
This makes it ineffective in isolation—but highly valuable when combined. 
These results establish a key principle: low-FPR detection relies more on calibrated signals.

\begin{figure}[t]
    \centering
    \includegraphics[width=0.9\linewidth]{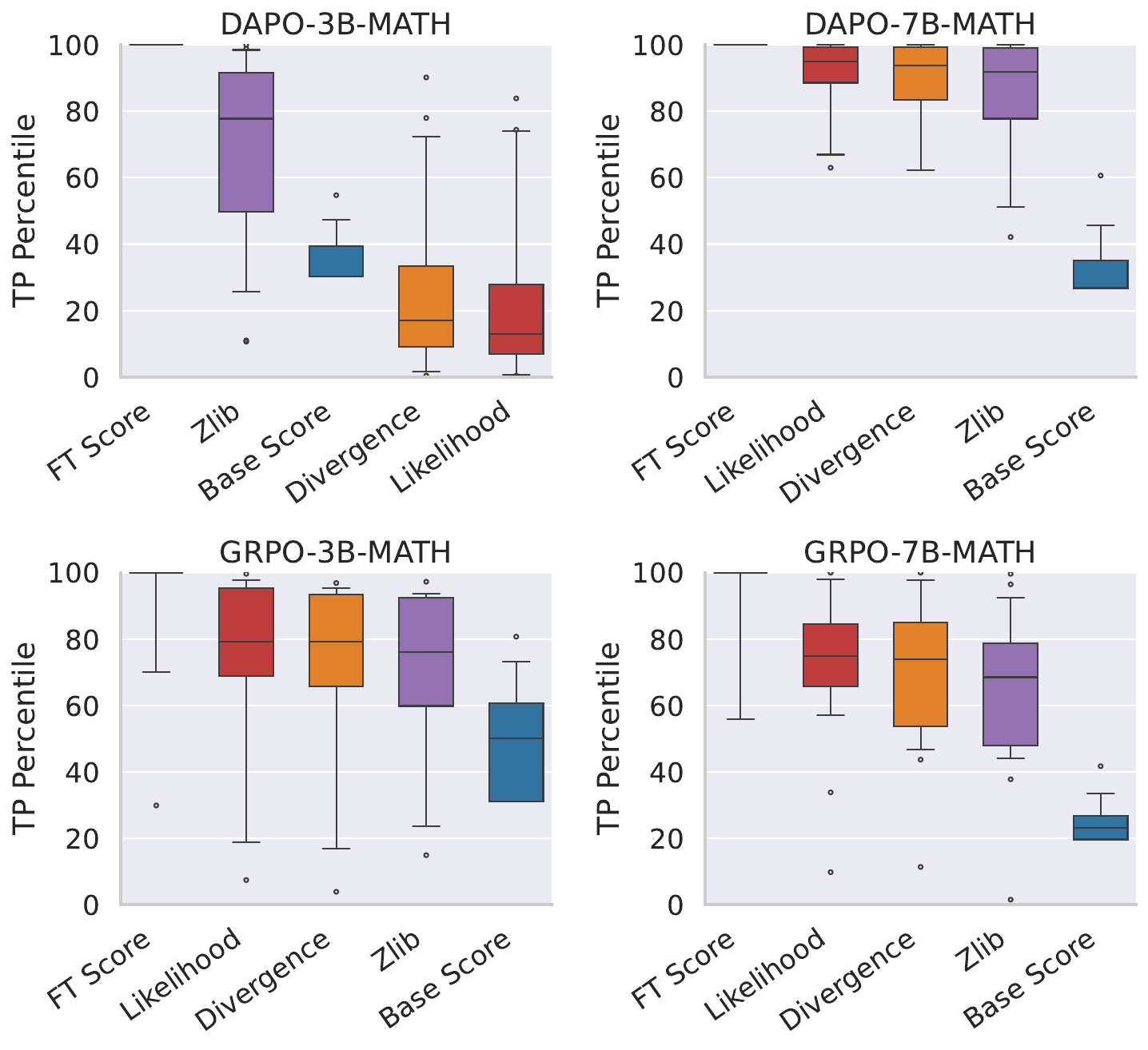}
    \caption{TP Percentile: Distribution of feature percentile ranks (vs. negatives) for true positives detected at 0.1\% FPR, across four runs.}
    \label{fig:hard-behave}
\end{figure}

\mypara{Ideal Behavior of TPs}
To understand how DIBA succeeds on clear true positives (TPs), those correctly identified at an extremely low FPR of 0.1\%, we examine how each feature of these TPs compares to the negative class distribution.
For each TP, we compute the percentile rank of its feature value within the negative distribution (see \Cref{fig:hard-behave}).

On the advantage side, clear TPs show a significant increase from Base Score to FT Score, indicating that fine-tuning meaningfully improved correctness on these prompts. 
This large improvement serves as a strong signal of training exposure.   
On the logit side, these instances consistently display high divergence and likelihood values, typically ranking above the 80th percentile of the non-member (negative) distribution. 
This indicates substantial token-level shifts in policy behavior, providing calibrated, high-confidence evidence for membership. 
The only exception is the DAPO-3B model, where TP feature values fall into low percentiles relative to negatives. 
Despite the counterintuitive direction, the separation in percentile space is extreme and highly consistent, enabling confident detection. 
These patterns can be further exploited for future work to guide data augmentation or feature engineering.
Additionally, we conduct an ablation study on removing one feature at a time, and the results are shown in \Cref{app:importance}.

\subsection{Error Analysis: When Does DIBA Fail?}
\label{sec:hardness}
Despite the intrinsic difficulty of the auditing caused by model capability~\Cref{sec:regimes}, we further analyze false positives to understand when DIBA may over-interpret exposure.

\begin{figure}[t] 
\centering 
\includegraphics[width=0.9\linewidth]{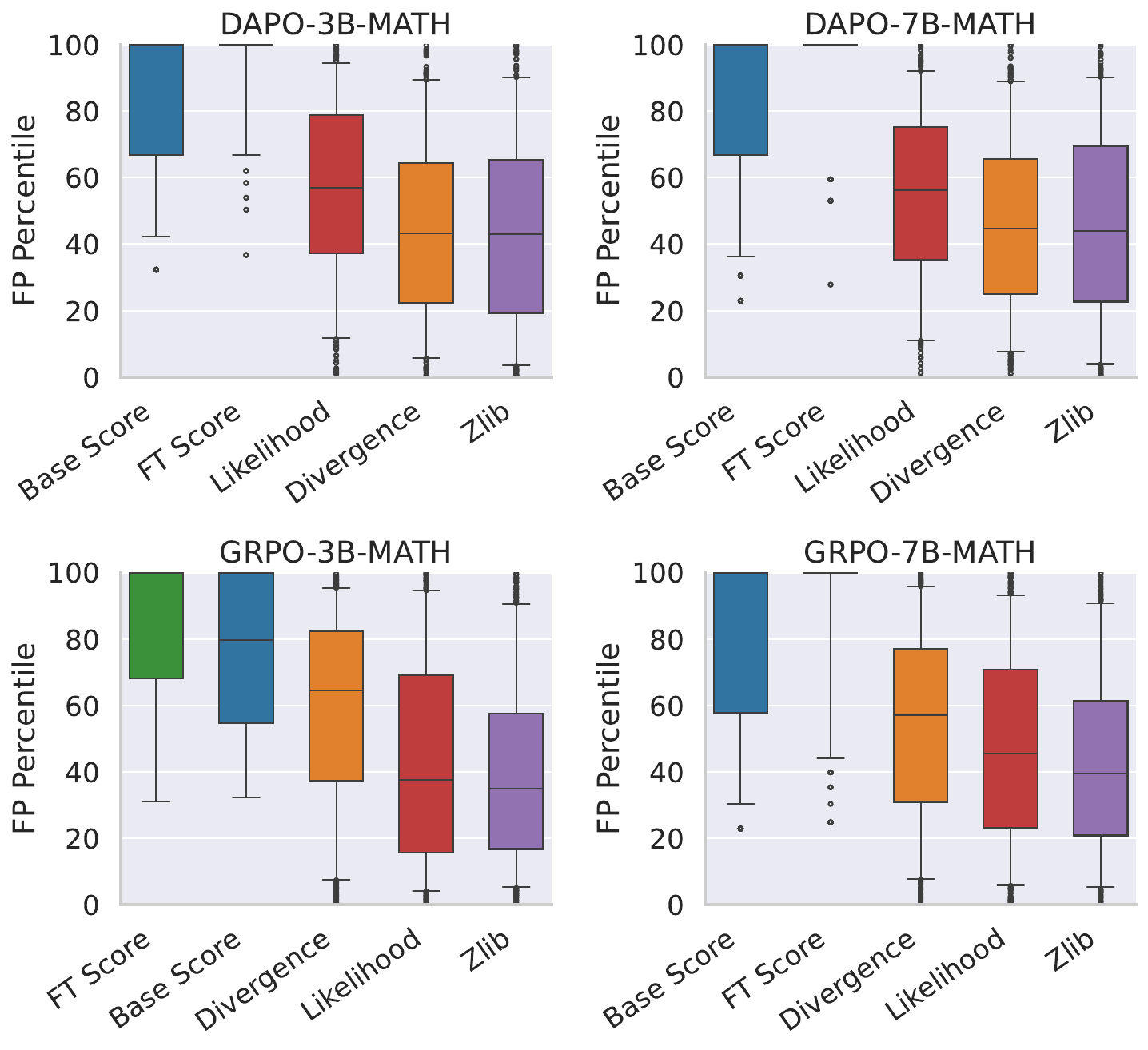} 
\caption{False Positive Percentiles: Distribution of feature percentile ranks for all false positives.} \label{fig:fp} 
\end{figure}

As shown in \Cref{fig:fp}, false positives consistently exhibit very high FT Score together with relatively high Base Score.
These are typically easy prompts that the base model already handles well, and whose performance is further reinforced by RLVR-like generalization.
As a result, their reward-side profile resembles that of genuine exposed prompts, even though they are not members.
At the same time, these false positives show only moderate or weak policy-side evidence: their divergence and likelihood-related features typically remain much less extreme than their reward-side scores.

Consequently, the policy-side signal is often insufficient to offset the strong reward-side evidence, leading to failure on easy prompts.
This reveals a central boundary of our method, that false positives arise when RLVR amplifies pre-existing capability on easy prompts.
In such cases, observable behavioral improvement is not unique to direct training inclusion, because non-member prompts may benefit indirectly through generalization.

Overall, these results sharpen the scope of RLVR auditing.
DIBA is most reliable on prompts that remain actively shaped by RLVR, but becomes ambiguous when training mainly reinforces capabilities the reference model already possesses.
\begin{figure*}[t]
    \centering
    \includegraphics[width=0.9\linewidth]{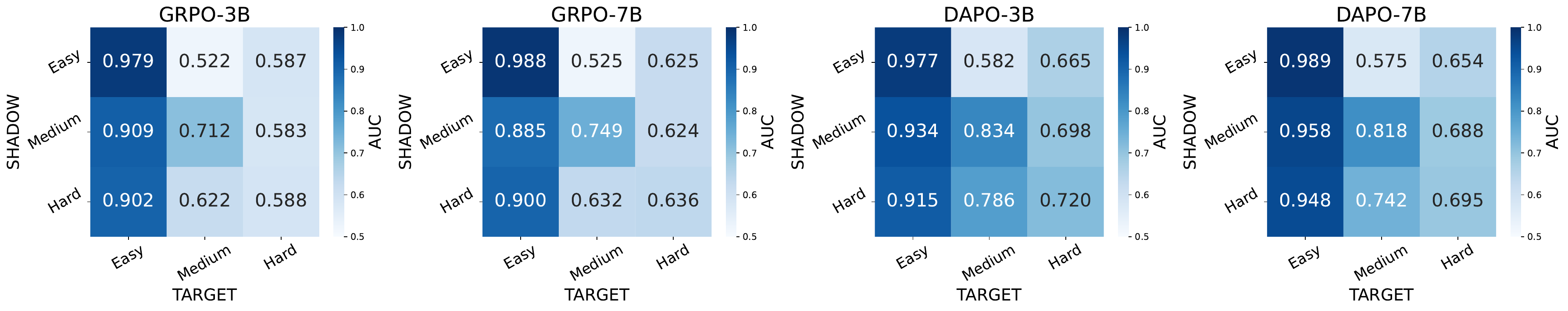}
    \includegraphics[width=0.9\linewidth]{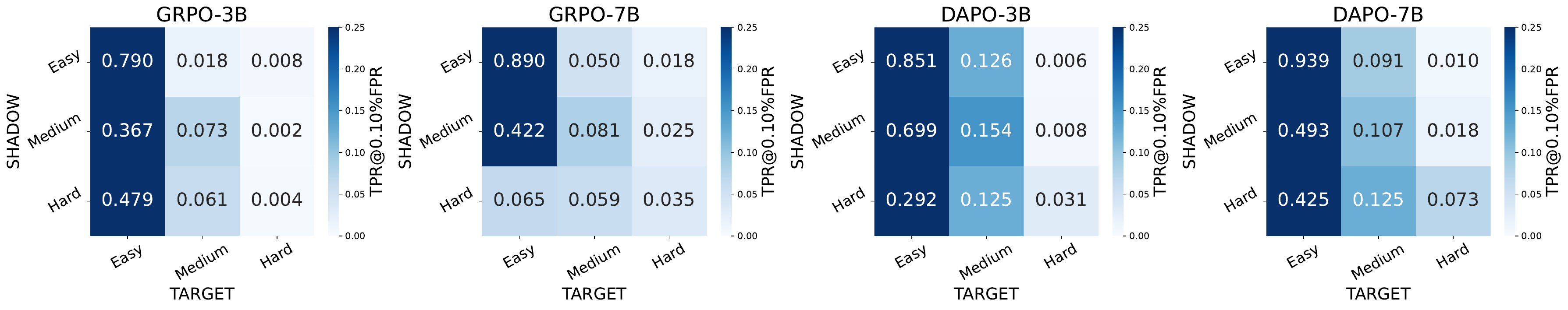}

\caption{Transferability across shadow datasets with different generalization regimes. Here, \textit{high-generalization} means harder for auditing because member-specific traces are less separable, while \textit{low-generalization} means easier for auditing.}
\label{fig:shadow-auc}
\end{figure*}

\subsection{Shadow Training}
\label{sec:shadow}

In this section, we study a practical grey-box setting in which the auditor knows the training algorithm and model architecture, but has no access to labeled data drawn from the same distribution as the target dataset.
As suggested by the fitting dynamics in \Cref{sec:fit}, transferability in our auditing problem is strongly mediated by how much prompt-specific improvement generalizes from members to non-members.
Accordingly, we organize shadow datasets by their \emph{generalization behavior} in our auditing setup, rather than by superficial task difficulty alone:
\begin{itemize}[leftmargin=*]
    \item \textit{Low-generalization set (easier for auditing):} improvement on members transfers weakly to non-members, producing a larger reward gap and stronger separability.
    \item \textit{High-generalization set (harder for auditing):} improvement transfers more broadly, resulting in similar behavior between members and non-members, leading to weaker separability.
\end{itemize}
This terminology is specific to our auditing problem, where ``high generalization'' means that member-specific traces are less distinguishable, not that the underlying task is intrinsically easier or better solved.

\mypara{Shadow Data Construction}
Due to the high computational cost of RLVR training, we train only a \emph{single} shadow model for this analysis.
Therefore, the results in this section should be interpreted as an exploratory study of transferability rather than a comprehensive characterization.
Detailed subset definitions and statistics are provided in \Cref{app:shadow}.
The data used in our main evaluation is categorized as a \textit{medium-generalization} target.
Additional in-distribution results in \Cref{app:shadow} confirm that the constructed subsets induce different degrees of auditability, consistent with different levels of member/non-member separation.

\mypara{Transferability Across Shadow Regimes}
We show the transferability results are shown in \Cref{fig:shadow-auc} and observe a consistent asymmetry:
\begin{itemize}[leftmargin=*]
    \item \textit{High-to-Low transfer:} detectors trained on the \textit{high-generalization} shadow set (harder for auditing, with subtler signals) transfer relatively well to \textit{low-generalization} targets (easier for auditing, with stronger signals).
    In our setting, these detectors maintain competitive AUC and better robustness in the low-FPR regime.
    This suggests that detectors trained to recognize weaker behavioral traces can often adapt to targets with stronger separation.
    
    \item \textit{Low-to-High transfer:} detectors trained on the \textit{low-generalization} shadow set transfer much less reliably to \textit{high-generalization} targets.
    When trained on easy-to-separate data, the detector tends to rely on large reward gaps and correspondingly degrades when transferred to targets where the behavioral signal is weaker and more nuanced.
\end{itemize}

Overall, these results suggest that shadow-data transfer in RLVR is sensitive to the strength and style of the underlying exposure signal.
When the target model's generalization properties are unknown, training on a harder-to-audit shadow distribution may provide a more conservative transfer strategy in our evaluated setting.
At the same time, this conclusion should be treated cautiously given the single-shadow design and the limited range of datasets considered here.
\begin{figure}[t]
    \centering
    \includegraphics[width=0.45\linewidth]{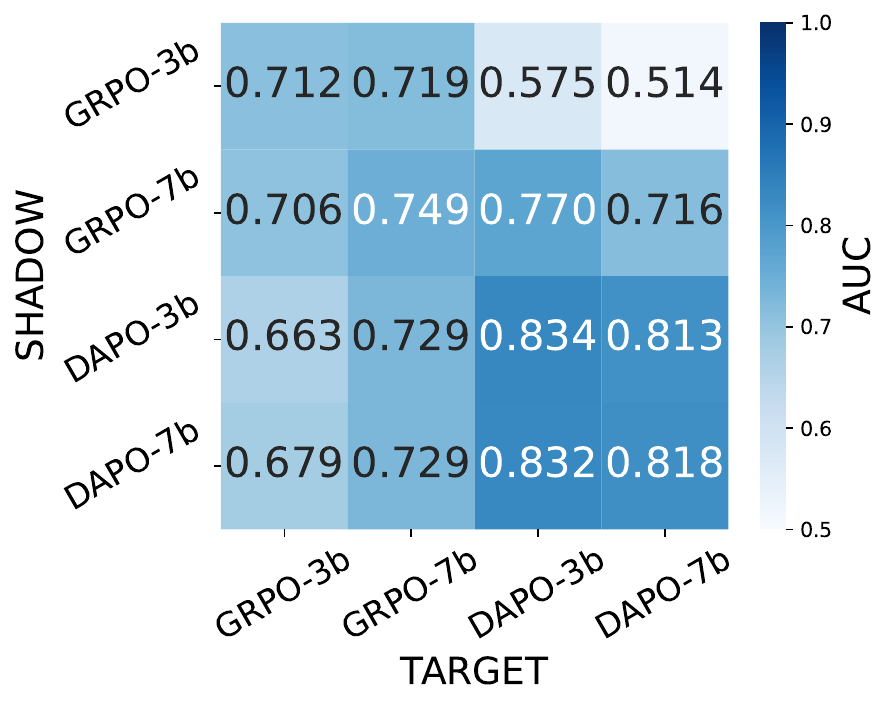}
    \includegraphics[width=0.45\linewidth]{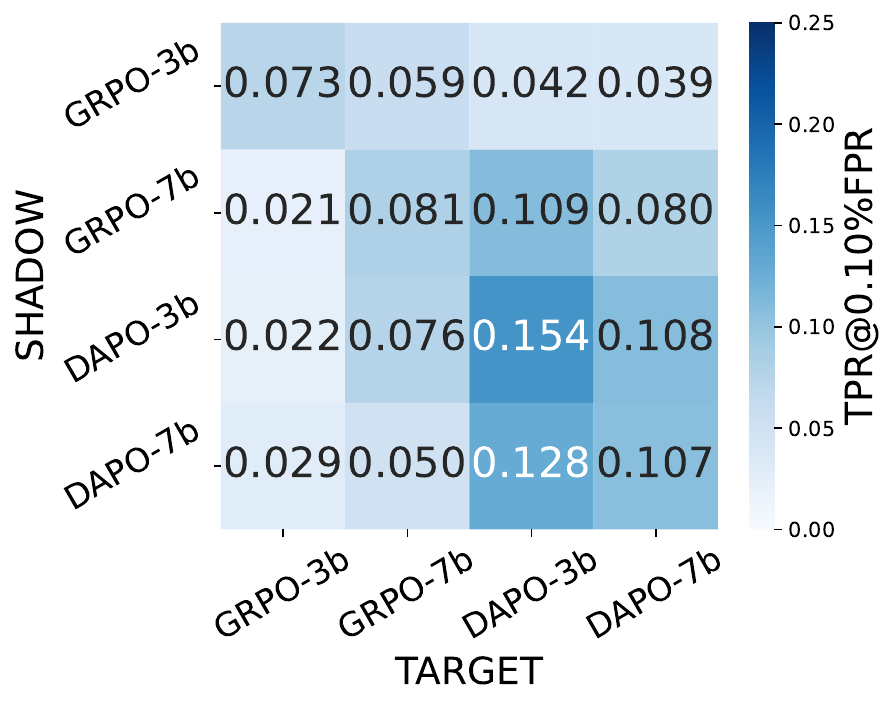}
\caption{Sensitivity of DIBA to base-model and RLVR-algorithm mismatch.}  
\label{fig:model-transfer}
\end{figure}

\subsection{Sensitivity to Base Model and RLVR Algorithm}
\label{sec:transfer}
We next study a practical grey-box setting in which the auditor does not know the target model's exact training recipe.
In particular, we test whether a DIBA detector trained on one RLVR configuration transfers across \emph{base-model scale} (3B vs.\ 7B) and \emph{RLVR algorithm} (GRPO vs.\ DAPO).
This should be viewed as a sensitivity analysis of the auditing signal under model and algorithm mismatch, rather than as a fully in-distribution evaluation.

The results in \Cref{fig:model-transfer} show three consistent patterns.
First, the matched setting, where the detector is trained and evaluated on the same model family and RLVR algorithm, achieves the strongest performance in most cases.
This is expected, since both the reward-side and policy-side traces are most consistent when the shadow and target configurations align.
Second, transfer across model sizes is relatively stable within the same RLVR algorithm.
For example, detectors trained on GRPO transfer reasonably between the 3B and 7B variants, and the same pattern largely holds for DAPO.
This suggests that changing scale alone does not dramatically alter the form of the exposure signal, at least within the same RLVR family.
Third, transfer across RLVR algorithms is substantially less reliable, especially in the high-precision regime.
When a detector trained on GRPO is applied to DAPO-finetuned models, or vice versa, performance drops much more sharply than in cross-size transfer, with the degradation being particularly clear in TPR@0.1\%FPR.
This result indicates that the observable exposure signal is sensitive to the training algorithm, and should not be assumed to transfer cleanly across different RLVR recipes.

Interestingly, detectors trained on DAPO often transfer better than detectors trained on GRPO.
One possible explanation is that DAPO induces a more consistent empirical policy-shift signal across the evaluated model scales.
However, we treat this as an empirical observation rather than a general claim, since our study only covers a limited set of base models and two RLVR algorithms.

Overall, these results refine the practical scope of DIBA.
The method transfers reasonably across model sizes within the same RLVR family, but is noticeably more sensitive to algorithm mismatch.
Therefore, when the target model's post-training recipe is unknown, DIBA should be interpreted as a useful but configuration-sensitive auditor rather than a recipe-agnostic compliance tool.

\begin{figure}[t]
    \centering
    \includegraphics[width=0.9\linewidth]{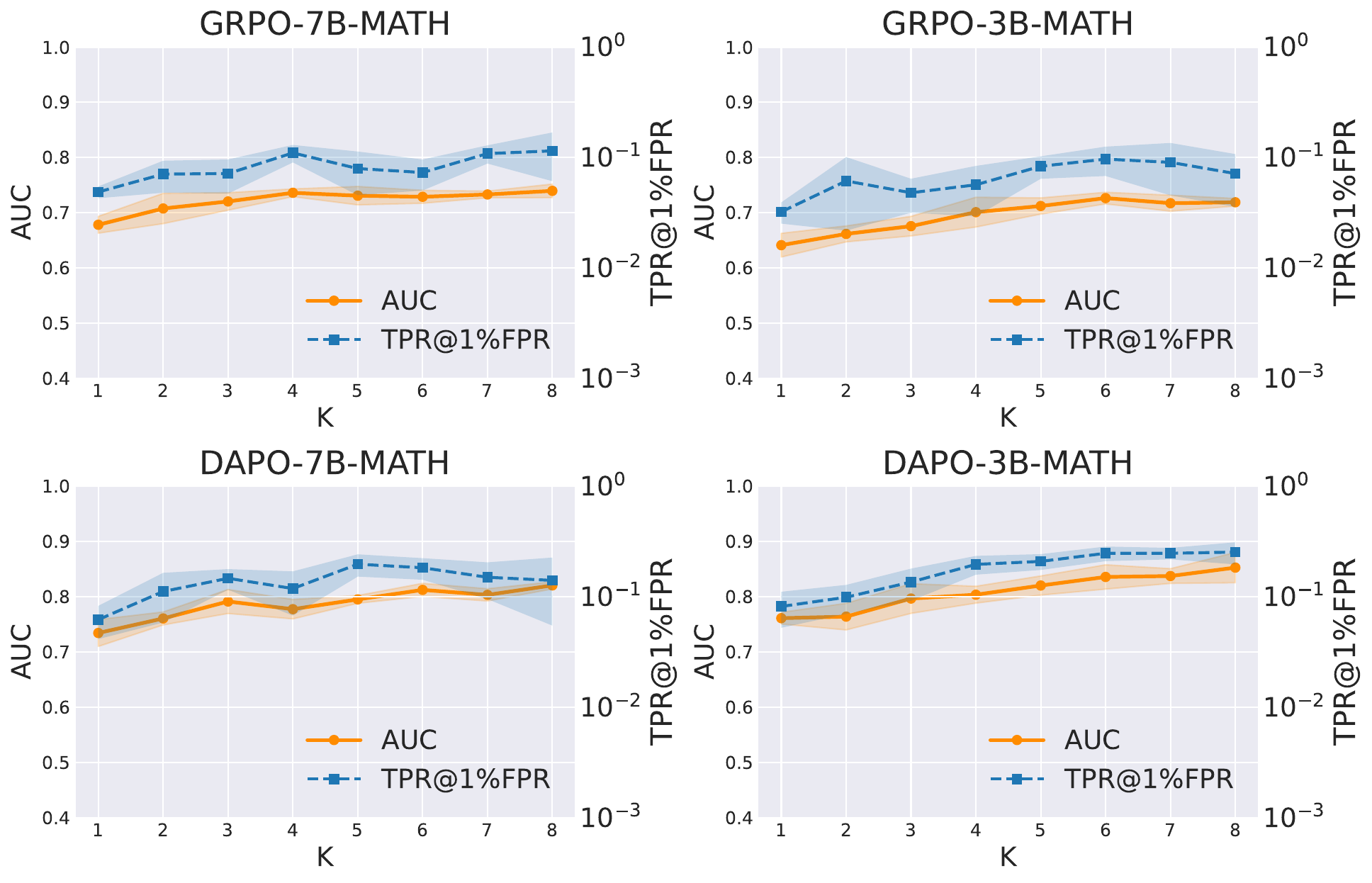}
    \caption{Ablation study on sampled responses.}
    \label{fig:ablate_sample}
\end{figure}

\subsection{Ablation Study}
\label{sec:ablate}
In this section, we conduct an ablation study to understand how the specific settings in DIBA may affect the final exposure auditing performance.

\mypara{Generated Samples}
Since DIBA relies on sampling multiple responses to compute both advantage and logit-level features, we conduct an ablation study on the number of generated responses per prompt (k). 
Ablation results in \Cref{fig:ablate_sample} are averaged over five independent runs, with shaded regions indicating the full range from minimum to maximum performance. 
When k is small, attack performance is suboptimal, as few samples inadequately capture the model’s true behavioral distribution and problem-solving ability. 
We observe that AUC saturates quickly, showing diminishing returns beyond k=4, while TPR@1\%FPR continues to improve with larger k, indicating that higher sample counts enhance the precision of outlier detection in the low-FPR regime by better characterizing tail behaviors. 
As a result, we set k=8 for all experiments in this paper.

\begin{figure}[t]
    \centering
    \includegraphics[width=0.9\linewidth]{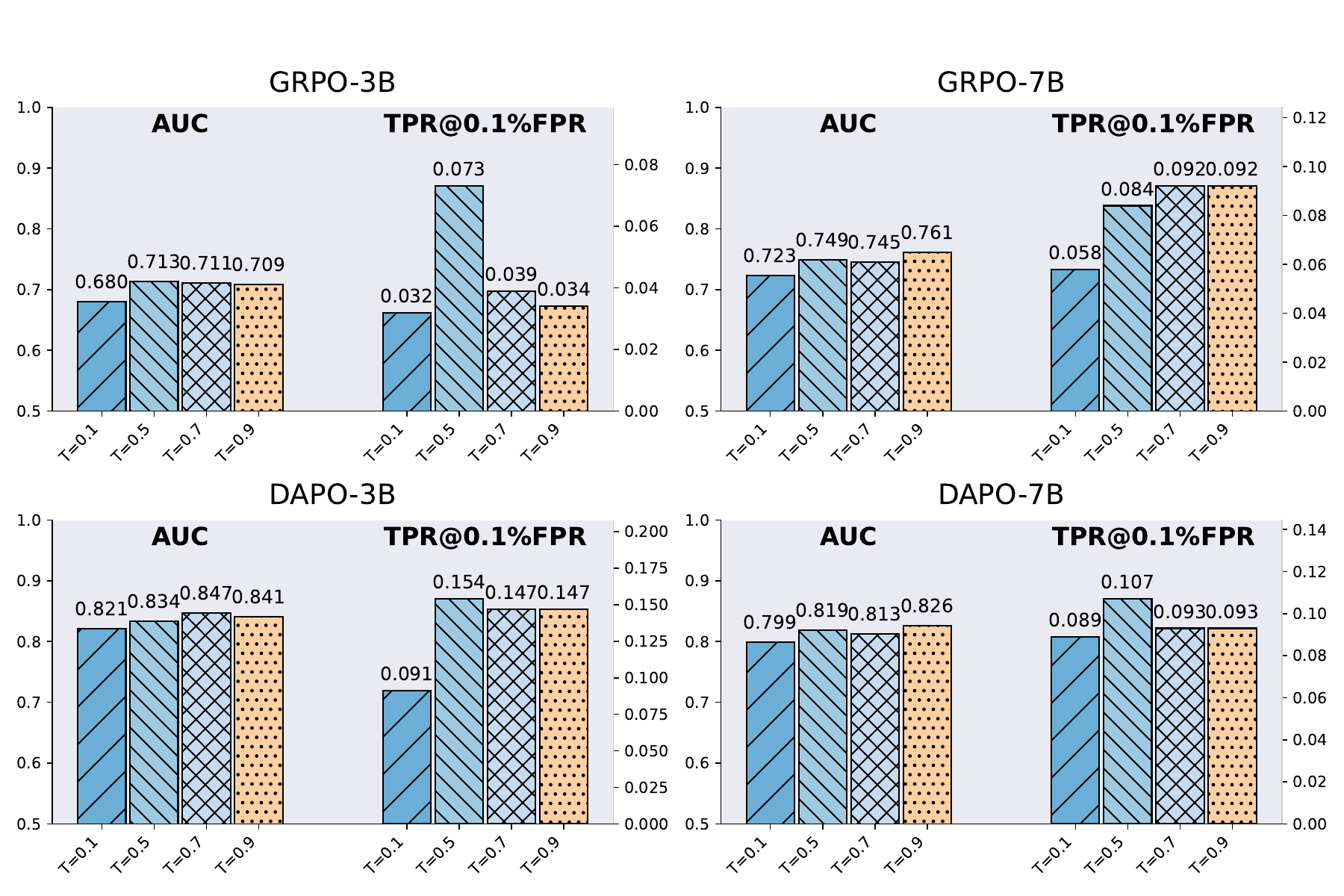}
    \caption{Ablation study on sampling temperatures.}
    \label{fig:ablate_temp}
\end{figure}

\mypara{Sampling Temperatures}
Another key hyperparameter in DIBA is the sampling temperature.
Since feature estimation relies on representative response distributions, the choice of temperature directly impacts the fidelity of behavioral change measurement. 
Results in \Cref{fig:ablate_temp} show the AUC and TPR@0.1\%FPR across different temperatures. 
We observe that temperature has only a modest effect on AUC: performance tends to be slightly lower at  low temperatures.

In contrast, TPR@0.1\%FPR is highly sensitive to temperature, with performance varying by up to a factor of two across settings. 
This suggests that tail behavior is critical for high-precision detection. 
At very high temperatures, excessive noise may obscure consistent policy shifts, while at very low temperatures, lack of variation limits the ability to detect subtle but systematic changes. 
Balancing robustness in both overall discrimination and high-precision detection, we find that a moderate temperature range (0.5–0.7) achieves optimal trade-offs. 
In this paper, we use $T=0.5$ to ensure sufficient diversity while maintaining coherent and verifiable outputs. 

\mypara{Response Source}
When calculating the logit side features, we base the calculation on responses sampled from the fine-tuned model. 
To assess the impact of this choice, we evaluate an alternative variant in which responses are instead sampled from the base model. 
Results, shown in \Cref{tab:ablate_response}, indicate that using fine-tuned model responses consistently yields higher TPR@0.1\%FPR across settings. 
Therefore, in this work, we adopt responses from the fine-tuned model to compute all logit-side divergence features. 

\subsection{Adaptation to VLMs}
\label{sec:vlm}

In this section, we verify whether the behavioral divergence mechanism extends beyond text-only models to Vision-Language Models (VLMs), verifying the modality-agnostic nature of DIBA.

\mypara{Setup}
We use Qwen-2.5-7B-VL-Instruct~\cite{bai2025qwen2} as the base model and fine-tune it on the Geo-3k dataset~\cite{lu2021inter}, which involves geometry problem-solving with visual inputs.
Crucially, our threat model remains focused on the {textual output distribution}: we measure the divergence in the generated reasoning path conditioned on the joint (Image, Text) prompt.
This validates whether visual context alters the detectability of policy optimization traces.
Implementation details are provided in \Cref{app:vlms}.

\begin{figure}[t]
    \centering
    \includegraphics[width=0.9\linewidth]{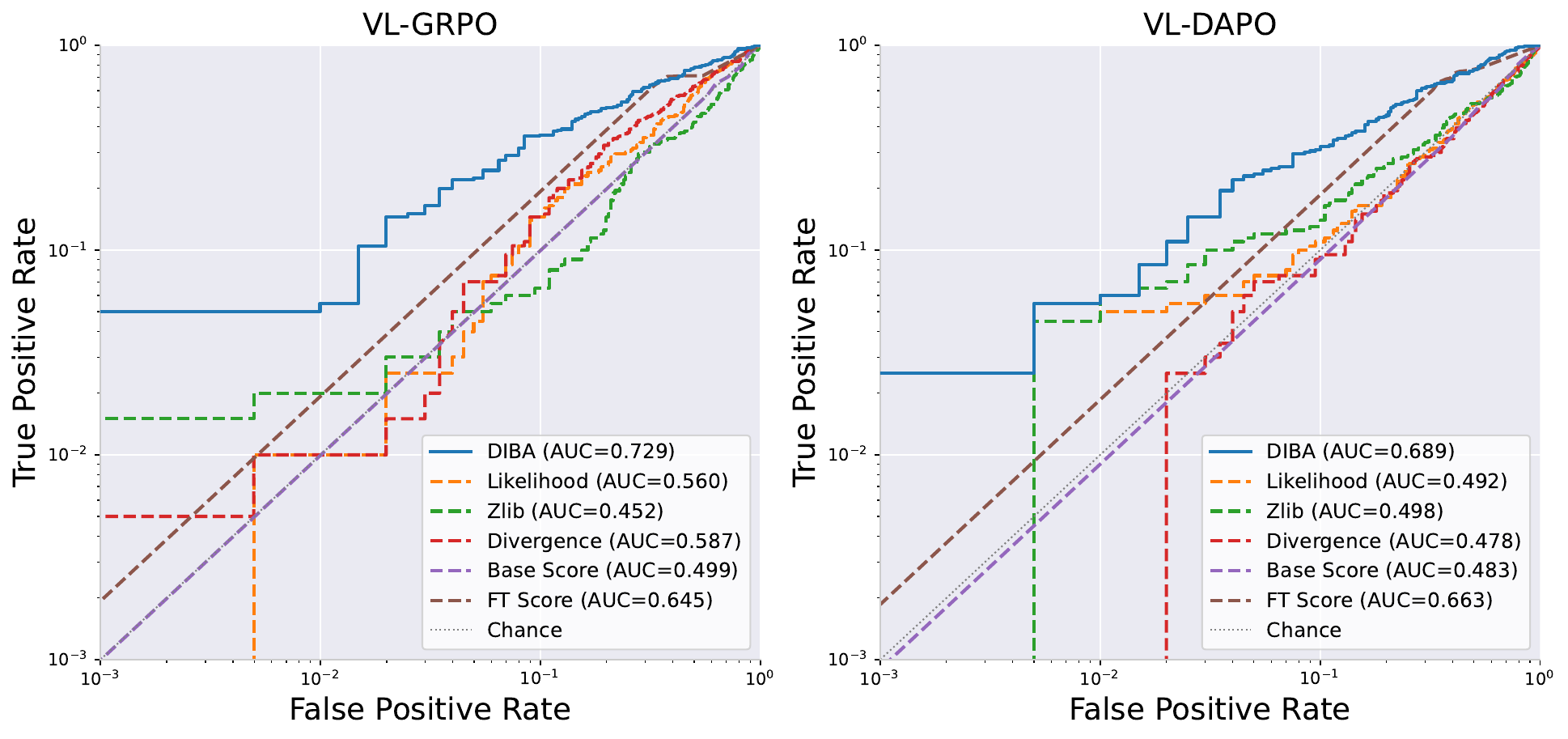}
    \caption{AUC and Log-scale ROC curves on VLMs. While AUC remains robust, low-FPR detection is challenging due to multimodal complexity.}
    \label{fig:roc_main_vl}
\end{figure}

\mypara{Results and Analysis}
Attack performance is shown in \Cref{fig:roc_main_vl} and \Cref{tab:main-vl} (in Appendix).
We observe that DIBA achieves a competitive AUC of $\approx$ 0.7, comparable to language-only settings. 
This confirms that RLVR induces measurable behavioral shifts even in multimodal latent spaces.
However, we identify two critical distinctions driven by modality:
\begin{itemize}[leftmargin=*]
    \item \textit{Dilution of High-Precision Signals.}
    While global separability (AUC) holds, the performance in the privacy-sensitive regime (TPR@0.1\%FPR) drops from the $10^{-1}$ range (LLMs) to the $10^{-2}$ range.
    We attribute this to the complexity of cross-modal alignment.
    Unlike pure text models where logit shifts are directly tied to semantic planning, VLM logits are conditioned on visual encodings. 
    The added variance from the visual encoder acts as a "noise floor," obscuring the subtle, tail-end probability shifts required for high-precision auditing.
    While the \texttt{FT Score} (correctness) remains a robust signal, the \texttt{Divergence} metric struggles to isolate policy drift from cross-modal noise.
    \item \textit{No-Learning, No-Leakage.}
    Contrary to the LLM setting, DAPO does not exhibit higher leakage than GRPO on VLMs.
    Upon analyzing the training logs (\Cref{app:vlms}), we find that DAPO suffered from suboptimal convergence on this dataset (likely due to the instability and immaturity of training framework).
    Far from being a limitation of DIBA, this result reinforces our core hypothesis regarding effective training:
    since the model failed to robustly internalize the training data (i.e., minimal gradient updates), there was no behavioral scar to detect.
    This unintended ablation study confirms that DIBA is not detecting artifacts of the algorithm, but the actual magnitude of information uptake.
\end{itemize}

\section{Reducing Observable Audit Signals}
\label{sec:defense}

We next examine whether the auditing signal exploited by DIBA can be reduced in practice.
Because our main setting assumes white-box models with access to logits, not all interventions have the same security meaning.
We therefore distinguish between (i) model-side or decoding-side interventions applied to the target model itself, and (ii) post-hoc output obfuscation applied only to released text outputs.
The former is relevant to our main threat model, whereas the latter should be interpreted as reducing \emph{text-level observability} rather than eliminating exposure in the underlying model.

\mypara{Model/Decoding-side Interventions}
Since DIBA compares the fine-tuned model against a reference checkpoint, a natural question is whether standard regularization or numerical perturbation can suppress the observable trace left by RLVR.
\begin{itemize}[leftmargin=*]
    \item \textit{Training Regularization.}
We first strengthen the KL regularization coefficient $\beta$ during RLVR training.
As shown in \Cref{tab:regularization}, stronger regularization has only a limited effect on DIBA in most evaluated cases: both AUC and TPR@0.1\%FPR remain broadly similar.
The main exception is Qwen-3B under GRPO, where TPR drops under large $\beta$.
However, this coincides with a collapse in training reward to near-base-model levels, suggesting that the apparent privacy gain comes from underfitting rather than genuine robustness.
Thus, increasing $\beta$ alone does not provide a practical defense once the model still learns meaningful task improvements.

\item \textit{Private Decoding and Noisy Logits.}
We further test perturbation at inference time, including private decoding and direct noise injection into logits.
According to \Cref{tab:dp-decode} and \Cref{fig:gau-auc}, both interventions only modestly reduce DIBA in our evaluated settings.
Even under strong perturbation, the decrease in AUC is limited, and high-precision auditing remains feasible in many cases.
Overall, these results suggest that the reward-side and policy-side traces used by DIBA are fairly robust to moderate numerical noise.

\end{itemize}
\begin{table}[t]
\centering
\caption{Comparison of post-hoc output obfuscation strategies. 
PPL is reported only as a fluency proxy, since answer-preserving paraphrasing leaves reward unchanged by construction in our math setting.}
\label{tab:defense_combined}
\resizebox{\linewidth}{!}{
\begin{tabular}{c|c|ccc|ccc} \toprule
 &  & \multicolumn{3}{c|}{GRPO} & \multicolumn{3}{c}{DAPO} \\ \cmidrule{3-8} 
\multirow{-2}{*}{Model} & \multirow{-2}{*}{Defense Strategy} & AUC & {\color[HTML]{000000} TPR} & PPL & AUC & {\color[HTML]{000000} TPR} & PPL \\ \midrule
 & No Defense (Original) & 0.712 & 0.073 & 1.09 & 0.834 & 0.154 & 1.15 \\
 & Char. Perturb ($\alpha=0.05$) & 0.662 & 0.027 & 2.26 & 0.789 & 0.093 & 2.53 \\
 & Char. Perturb ($\alpha=0.10$) & 0.635 & 0.013 & 3.35 & 0.768 & 0.076 & 3.80 \\
\multirow{-4}{*}{Math-3b} & \textbf{Paraphrasing (Llama-3)} & \textbf{0.545} & \textbf{0.003} & \textit{--} & \textbf{0.668} & \textbf{0.012} & \textit{--} \\ \midrule
 & No Defense (Original) & 0.749 & 0.084 & 1.13 & 0.818 & 0.107 & 1.19 \\
 & Char. Perturb ($\alpha=0.05$) & 0.710 & 0.017 & 2.27 & 0.756 & 0.067 & 2.41 \\
 & Char. Perturb ($\alpha=0.10$) & 0.679 & 0.008 & 3.33 & 0.728 & 0.055 & 3.55 \\
\multirow{-4}{*}{Math-7b} & \textbf{Paraphrasing (Llama-3)} & \textbf{0.587} & \textbf{0.012} & \textit{--} & \textbf{0.658} & \textbf{0.020} & \textit{--} \\\bottomrule
\end{tabular}
}
\end{table}

\mypara{Post-hoc Output Obfuscation}
We also study post-hoc transformations of generated text, which do not protect the model itself but only reduce text-level observability.
Character perturbation (see \Cref{app:defense}) lowers DIBA only at severe utility cost.
Paraphrasing using an external LLM (see \Cref{app:defense}) is more effective: when constrained to preserve the final answer, it weakens the link between the prompt and the original optimized output, reducing text-output-based auditability.
In our evaluated math setting, the verifier depends only on the final answer, so answer-preserving paraphrasing leaves task reward unchanged by construction; accordingly, we report perplexity only as a fluency proxy rather than re-evaluating task reward.
This benefit is limited to text-only settings and does not remove exposure under white-box auditing.

Overall, our results suggest a clear distinction:
model-side regularization and numerical perturbation provide limited protection against DIBA in the evaluated white-box setting, whereas paraphrasing can reduce the externally observable text-level auditing signal, but does not eliminate exposure under stronger auditors with access to the model or its logits.
\begin{figure*}[t]
    \centering
    \includegraphics[width=0.7\linewidth]{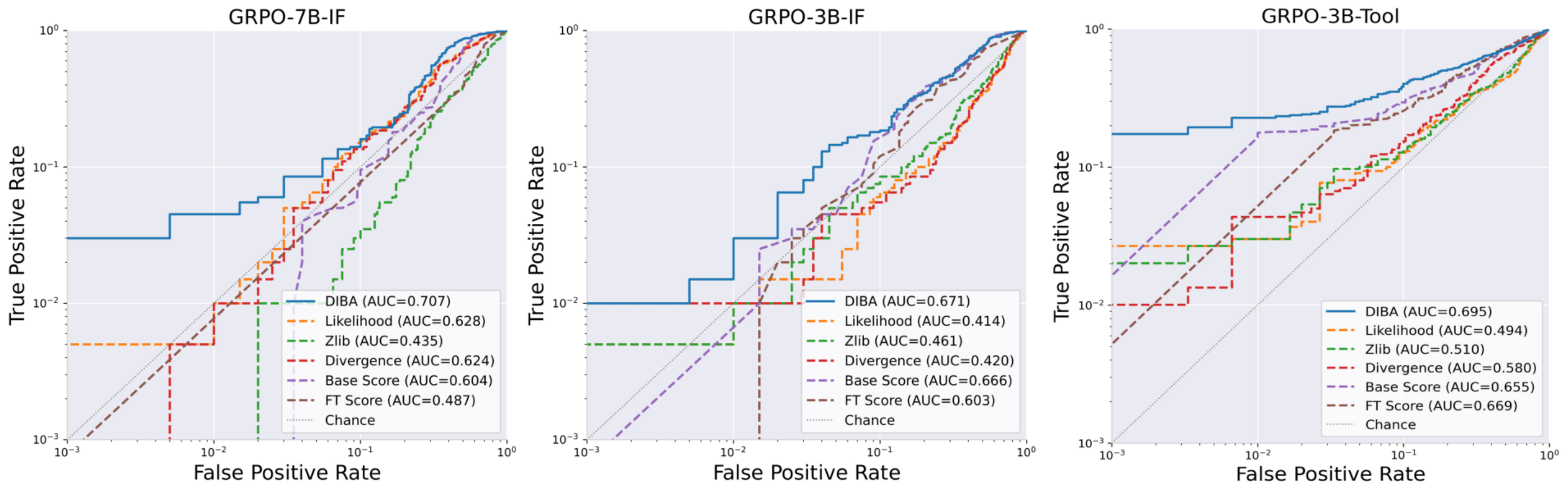}
    \caption{Adaptation DIBA to instruction-following and Tool-using dataset.}
    \label{fig:roc-if}
\end{figure*}

\begin{figure}[b]
    \centering
    \includegraphics[width=0.49\linewidth]{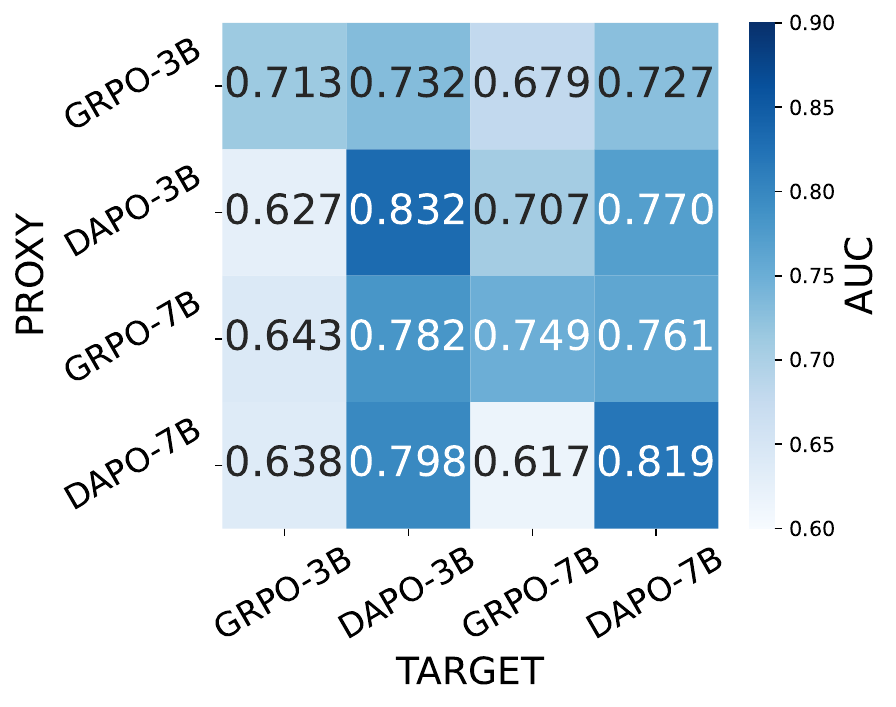}
    \includegraphics[width=0.49\linewidth]{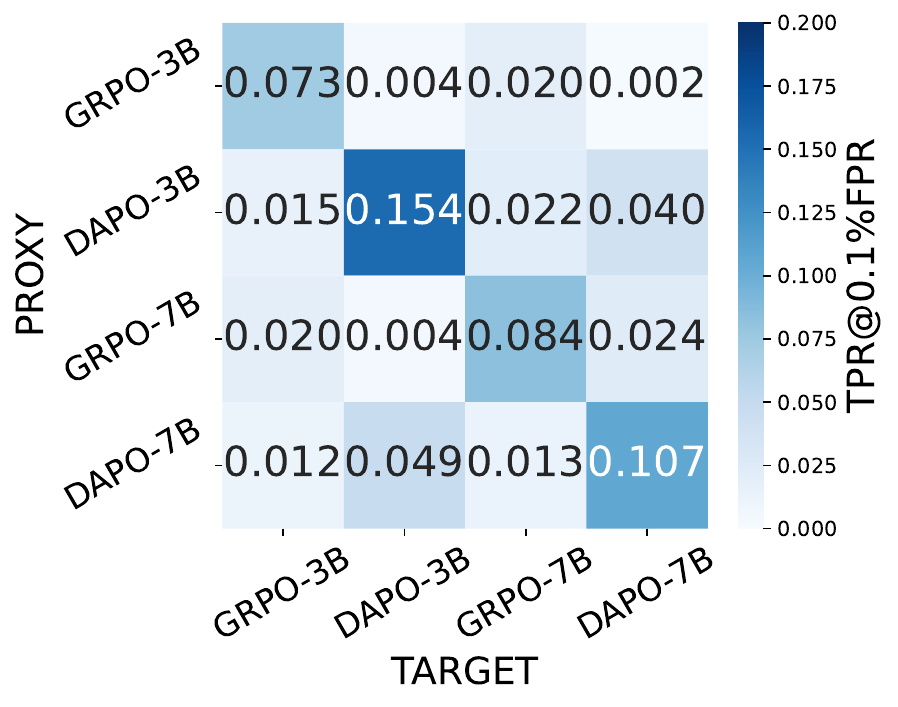}
    \caption{Black-box performance of DIBA.}
    \label{fig:balckeval}
\end{figure}

\section{Discussions and Limitations}
\label{sec:discussion}

\mypara{Black-box Feasibility vs. White-box Auditing}
We investigate the feasibility of applying DIBA in a strict black-box setting, where the auditor has no access to model weights or logits, relying solely on generated text.
As shown in~\Cref{fig:balckeval}, estimating logit-side features via proxy models yields mixed results.
While DAPO-trained proxies enable non-trivial detection (significantly above random), GRPO-trained proxies yield weak signals in the high-precision regime (TPR@0.1\%FPR $<$ 0.01).
This performance gap underscores a fundamental characteristic of DIBA: it is designed as a precision auditing tool, not a blind attack.
The behavioral divergence we target, i.e., subtle shifts in token probability distributions driven by policy gradients, is inherently difficult to reconstruct without logit access.
Therefore, while DIBA can function in black-box settings with matched proxies, its primary utility lies in the open-weight compliance setting, where access to logits allows for the definitive detection of training data exposure.

\mypara{Applicability Boundaries: Domains and Tasks}
Our evaluation focuses primarily on math-related datasets because the RLVR paradigm is intrinsically defined by objective, verifiable rewards.
Domains such as creative writing or subjective QA typically rely on preference models (RLHF/DPO) rather than rule-based verification, falling outside our threat model.
Regarding {Code Generation}, another major verifiable task, we excluded it from this study due to the technical maturity of current open-source RLVR frameworks. 
We observed that code RLVR often suffers from much sparser reward signals (pass/fail unit tests) compared to the dense verification possible in math, leading to unstable convergence that obscures distinct behavioral patterns.
However, to validate transferability beyond math, we successfully adapted DIBA to two additional verifiable domains:
(1)~\textit{Instruction Following}~\cite{ye2025multi}, where rewards are computed based on strict constraint adherence (see \Cref{app:if}), and
(2)~\textit{Tool Use}, where a GRPO-trained model learns structured function-calling via correctness rewards that verify argument accuracy against ground-truth tool invocations (see \Cref{app:tool}).
Results in \Cref{fig:roc-if} confirm that behavioral divergence is not limited to reasoning:
both Instruction Following (AUC $\approx$ 0.71) and Tool Use (AUC $\approx$ 0.70, TPR@0.1\%FPR $\approx$
23\%) exhibit strong membership signals, demonstrating that DIBA generalizes to any domain where policy optimization is driven by hard constraints.

\mypara{The ``Pre-existing Knowledge'' Limitation}
A critical limitation of our approach, and indeed of auditing RLVR in general, is the inability to detect leakage of ``known'' knowledge.
As analyzed in \Cref{sec:hardness}, DIBA relies on \textit{actual learning traces} (gradient updates).
If a prompt is already perfectly solved by the base model (the "All-1" subset), the reward signal is saturated, gradients approach zero, and no significant behavioral shift occurs.
Consequently, DIBA cannot determine whether such a prompt was present in the training set or simply ignored due to mastery.
This is an inherent physical boundary: auditability in RLVR is limited to data that actively contributes to model alignment, leaving pre-existing knowledge transparent to inference.

\mypara{Models and Algorithms}
We focus on GRPO and DAPO, the two representative RLVR algorithms, using the Qwen series (3B and 7B). 
While we expect our findings on behavioral divergence to generalize to other policy gradient methods (e.g., PPO) and architectures (e.g., Llama), we leave the exhaustive verification of these combinations to future work.

\section{Conclusion}
We study privacy auditing in RLVR through query-level exposure rather than static memorization.
We show that RLVR can leave observable prompt-specific traces in a released model, and propose DIBA to audit them through reward-side change and policy-side divergence.
Our evaluation also clarifies the boundary of this signal: DIBA is most effective on prompts that continue to induce optimization, and less reliable on already-mastered prompts.
Finally, we find that paraphrasing can reduce text-level auditability in output-only settings, but does not remove exposure under white-box or logit-access auditing.
These results position DIBA as a practical yet configuration-sensitive auditor for training-data exposure in RLVR.

\section*{Ethics Considerations}
This work highlights critical privacy risks in RLVR, a paradigm increasingly used to arm large language models with advanced reasoning capabilities.
While RLVR avoids explicit supervision and reference answers, our attack DIBA demonstrates that behavioral changes induced by training can still leak information.
This raises ethical concerns regarding data confidentiality, especially when training data contains sensitive, personal, or proprietary content.

We emphasize that this research is intended to improve understanding of privacy vulnerabilities and strengthen defenses, not to enable harmful exploitation. 
The release of any implementation will follow responsible disclosure practices, limited to academic use with appropriate safeguards. 
We hope this work encourages the community to develop more privacy-preserving RLVR methods and to consider behavioral leakage as a legitimate risk in future alignment efforts.

\section*{Open Science}
All artifacts necessary to evaluate and reproduce the results are provided as follows:

\mypara{Datasets}
We provide processed splits of the MATH and NuminaMath datasets used for the main evaluation (member and non-member sets).
We provide the Easy and Hard (synthetic) shadow datasets used for the transferability and generalization analysis.
We provide the Geo-3k visual reasoning dataset used for the VLM adaptation.
Due to the large size of the training checkpoint, we provide the training script that can be used in the VeRL framework.

\mypara{Baselines}
Implementations of standard memorization-based MIAs, including LOSS, LiRA, Zlib, and Min-K\%, adapted for the LLM setting.

\mypara{Inference Scripts}
The complete DIBA framework, including scripts for advantage-side reward estimation and logit-side K3 divergence computation.
Evaluation scripts for reporting metrics such as AUC, balanced accuracy, and TPR@0.1\%FPR.
The artifacts are available in \url{https://github.com/Y-L-LIU/Divergence-in-Behavior-Attack}.

\bibliographystyle{IEEEtran}
\bibliography{ref}

\appendix
\section{Related work}
\begin{itemize}[leftmargin=*]
    \item \textbf{LOSS~\cite{yeom2018privacy}:}
    The simplest baseline uses the negative log-likelihood (NLL) under the target model:
    \[
    \text{Score}_{\text{LOSS}}(q) = -\log \pi_\theta(q) = -\sum_{i=1}^{|q|} \log \pi_\theta(q_i \mid q_{<i}).
    \]
    A threshold $\tau$ is chosen (e.g., via validation data), and $q$ is classified as a member if $\text{Score}_{\text{LOSS}}(q) \leq \tau$.

    \item \textbf{LiRA~\cite{carlini2022membership}:}
     Compares the target model’s likelihood to that of a reference model $\pi_{\theta_{\text{ref}}}$ trained on disjoint data:
    \[
    \text{Score}_{\text{LiRA}}(q) = \log \pi_\theta(q) - \log \pi_{\theta_{\text{ref}}}(q),
    \]
    where higher scores indicate membership.

    \item \textbf{Zlib~\cite{carlini2021extracting}:}
    Normalizes the model loss by the text’s compressibility using the zlib algorithm. Let $\text{zlib}(q)$ be the compressed size (in bits) of $q$. Then:
    \[
    \text{Score}_{\text{zlib}}(q) = \frac{-\log \pi_\theta(q)}{\text{zlib}(q)},
    \]
    where lower values suggest membership.
    \item \textbf{Min-k\%:}
    Uses only the $k\%$ of tokens with the lowest per-token loss. Let $\mathcal{T}_k(q)$ be the set of indices of these tokens. Then:
    \[
    \text{Score}_{\text{Min-}k\%}(q) = -\frac{1}{|\mathcal{T}_k(q)|} \sum_{i \in \mathcal{T}_k(q)} \log \pi_\theta(q_i \mid q_{<i}),
    \]
    where higher scores (i.e., lower average loss on predictable tokens) indicate membership.
\end{itemize}

\section{Preliminary for RL}\label{app:preliminary}
\mypara{PPO~\cite{schulman2017proximal}}
It is one of the most widely used on-policy reinforcement learning algorithms in the context of LLM alignment~\cite{bai2022training}. 
To stabilize training, PPO assumes that the model only has a single update following each exploration stage.
Specifically, for a prompt $q$ and generated response $o = (o_1, \dots, o_T)$, PPO computes an advantage $A_t$ for each token $o_t$, which measures how much better or worse the token is compared to the average behavior under the current policy. 
The advantage is typically estimated using a separate value network, which predicts the expected future reward.
The PPO objective is:
\begin{equation}
{\begin{split}
\mathcal{J}(\theta) = \mathbb{E}_{[q, o \sim \pi_{\theta_{\text{old}}}]}  \sum_{t=1}^{|o|}  \frac{\pi_\theta(o_t \mid q, o_{<t}) }{\pi_{\theta_{\text{old}}}(o_t \mid q, o_{<t})}A_t,
\end{split}
}\end{equation}
A key drawback of PPO is that the training may involve an additional value network of comparable size, which doubles memory usage and complicates training.

\mypara{DAPO}
The optimization objective of DAPO is shown as:
\begin{equation}\label{eq:lossdapo}
\begin{split}
&\mathcal{J}(\theta) = \mathbb{E}_{[q, \{o_i\}_{i=1}^G \sim \pi_{\theta_{\text{old}}}]} \\
&\Bigg[
\underbrace{\frac{1}{\sum_{i=1}^G |o_i|} \sum_{i=1}^G \sum_{t=1}^{|o_i|}}_{\text{Rebalancing Act}}
\frac{\pi_\theta(o_{i,t} \mid q, o_{i,<t})}{\pi_{\theta_{\text{old}}}(o_{i,t} \mid q, o_{i,<t})}
\underbrace{\hat{A}_{i,t}^{\text{clipped}}}_{\text{Clip Higher}}
)+\underbrace{R_{l}(\{o_i\}_{i=1}^G)}_{\text{Overlong Punish}}
\Bigg], \\
& \text{s.t.} \quad \underbrace{0 < \big|\{o_i \mid \texttt{is\_correct}(o_i)\}\big| < G}_{\text{Dynamic Sampling}},
\end{split}
\end{equation}
where $\texttt{is\_correct}(o_i)$ represents the if the responses contains correct answer.
\section{Baseline Implementation}\label{app:base-imple}
For the neighbor baseline, we use RoBERTa-base~\cite{liu2019robertarobustlyoptimizedbert} to encode each prompt and retrieve the 8 nearest neighbors based on cosine similarity in the embedding space. 
For the min-k\% baseline, we set k=0.2.
To calibrate each method, we compute the relevant metric (e.g., likelihood, entropy) separately on responses generated by the base and fine-tuned models, then take the difference: 
$\text{metric}_{\text{ft}} - \text{metric}_{\text{ref}}$. 
As shown in prior work~\cite{ran2025lora}, this calibration effectively removes biases present in the pre-trained model and isolates changes due to fine-tuning. 
Such calibration can improve performance in low-FPR regimes.
For the self-traject setting, we first sample responses from the fine-tuned model and compute the calibrated metrics like before.
For prediction, we adopt a similar stack predictor in our main evaluation.

\section{Hyperparameters}\label{app:hyper}
\mypara{RLVR Setup}
Both GRPO and DAPO use a training batch size of 512 prompts and generate 16 responses per prompt. 
They share the same learning rate (1e-6), max prompt length (GRPO: 1024; DAPO: 2048), and max response length (GRPO: 2048; DAPO: 8192). 
Both employ PPO mode in verl with a mini-batch size of 512 (GRPO) or 32 (DAPO). 
The micro batch size on each GPU is set to 8 for both algorithms to avoid out-of-memory errors.
KL loss and reward KL are turned off in DAPO; while GRPO uses KL loss with a coefficient of 0.001.
DAPO introduces asymmetric PPO clipping (clip ratios [0.2, 0.28]) and uses token-mean loss aggregation, while GRPO uses standard symmetric clipping. 
Both run for 60 epochs on 4 GPUs per node with vLLM-based rollout and gradient checkpointing enabled.

\mypara{Implementation Details}
During response generation, we sample 8 distinct responses per prompt using temperature 0.5, top-p sampling with $p=0.9$, and a repetition penalty of 1.05  to balance diversity and coherence. 
For each response, we compute sample-level features, which are then aggregated at the prompt level. 

For the predictor, we utilize a stacking classifier that combines a Random Forest (with 100 trees) and a Logistic Regression pipeline (which includes median imputation with missing indicators and column-wise standard scaling) as base estimators to generate meta-features.
The final estimator is a Logistic Regression model trained on these meta-features to produce the final prediction.

\section{Feature Importance}\label{app:importance}
To understand the effect of each feature component, we conduct an ablation study by removing one feature at a time.
We find that sequentially removing the logit-side feature degrades the TPR@0.1\%FPR performance.
When only the \texttt{Ft Score} is left, the metrics drops to zero.

\begin{table}[ht]
\centering
\label{tab:importance}
\caption{Ablation study on removing component features of DIBA.}
\begin{tabular}{l r}
\toprule
\textbf{Ablation Setting} & \textbf{TPR@0.1\%FPR} \\
\midrule
Baseline (all features) & 15.4\% \\
\quad - Remove \texttt{Zlib} & 15.3\% \\
\quad - Remove \texttt{Divergence} & 14.6\% \\
\quad - Remove \texttt{Likelihood} & 11.4\% \\
\quad - Remove \texttt{Base Score} & 0.0\% \\
\bottomrule
\end{tabular}
\end{table}
\section{Hard Example}
\begin{figure}[h]
    \centering
    \includegraphics[width=1\linewidth]{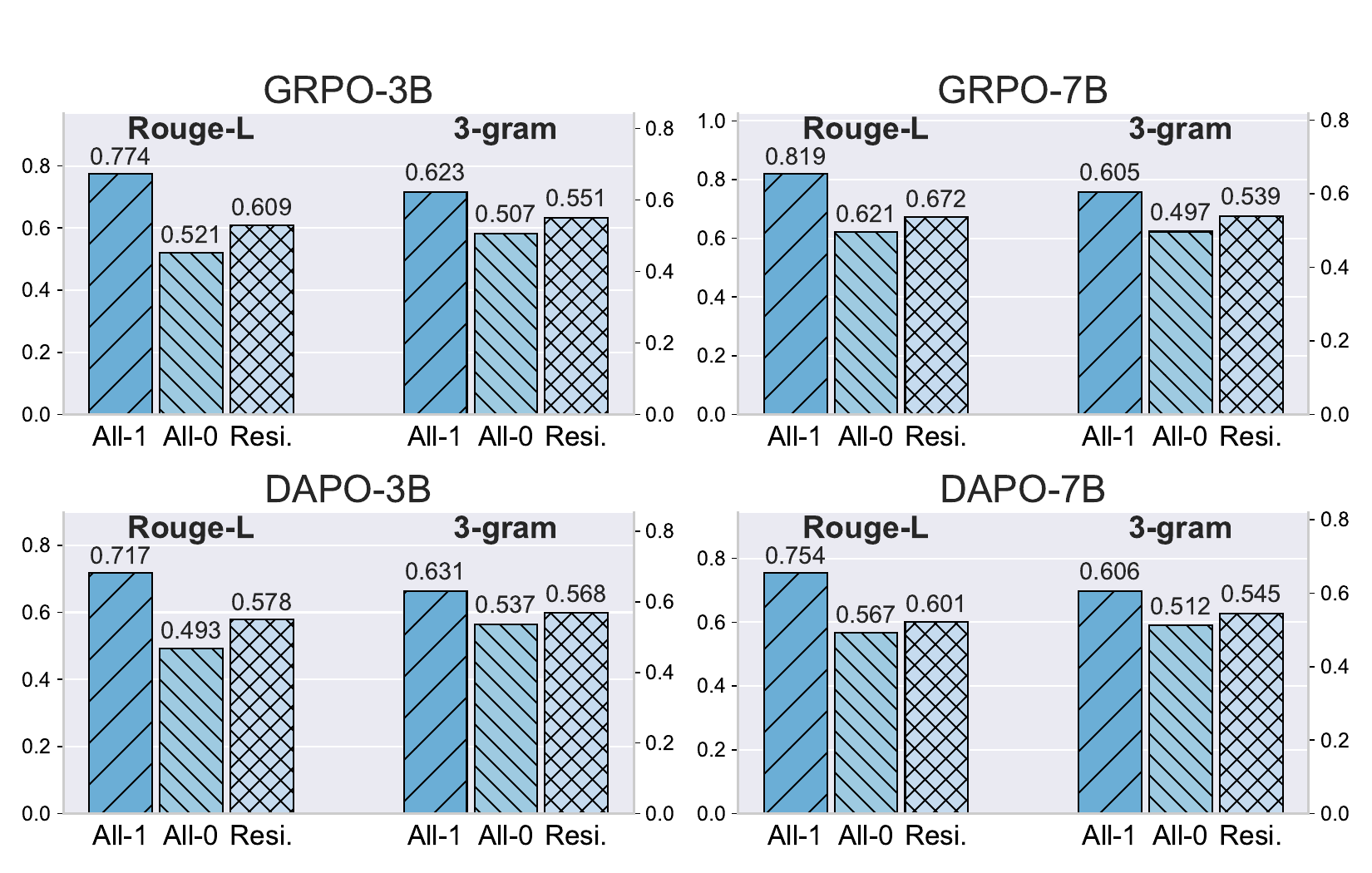}
    \caption{Similarity Between Responses Generated by Fine-tuned Model and Base Model: For each prompt, we generate 8 responses, and compute the group Rouge-L and 3-gram to get the prompt-level similarity, then average over the whole dataset (both member and non-member data).   }
    \label{fig:hard_sim}
\end{figure}

See \Cref{fig:hard_sim}.
\section{Ablation Study}
See \Cref{tab:ablate_response}.
 \begin{table}[t]
\centering
\caption{Ablation study on response sources.}
\label{tab:ablate_response}
\large
\resizebox{0.95\linewidth}{!}{\begin{tabular}{c|c|cc|cc}\toprule
 &  & \multicolumn{2}{c|}{GRPO} & \multicolumn{2}{c}{DAPO} \\ \cmidrule{3-6} 
\multirow{-2}{*}{Model} & \multirow{-2}{*}{Response} & AUC & {\color[HTML]{000000} TPR@0.1\%FPR} & AUC & {\color[HTML]{000000} TPR@0.1\%FPR} \\ \midrule
 & Base & 0.704 & 0.036 & 0.846 & 0.113 \\
\multirow{-2}{*}{Math-3b} & Fine-tuned & 0.712 & 0.073 & 0.834 & 0.154 \\ \midrule
 & Base & 0.764 & 0.076 & 0.809 & 0.113 \\
\multirow{-2}{*}{Math-7b} & Fine-tuned & 0.749 & 0.084 & 0.818 & 0.107 \\\bottomrule
\end{tabular}
}
\end{table}

\section{Shadow Training}\label{app:shadow}
\mypara{Dataset}
For the hard dataset, we use a synthetic variant of MATH~\cite{numina_math_datasets} in which prompts and solutions are systematically paraphrased to increase linguistic and structural complexity, making generalization more challenging. 
For the easy dataset, we sample directly from the original MATH training set, preserving its natural formulation and higher learnability. 
From both sets, we select 1,000 members and 1,000 non-member prompts. 
The dataset in the main evaluation is treated as the medium dataset.

\begin{figure*}
    \centering
    \includegraphics[width=0.9\linewidth]{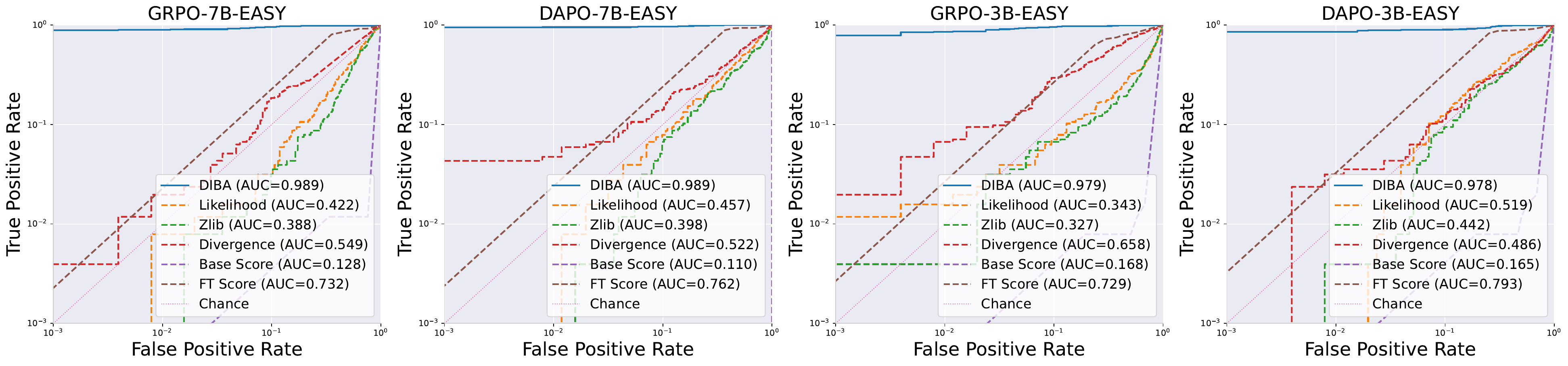}
    \includegraphics[width=0.9\linewidth]{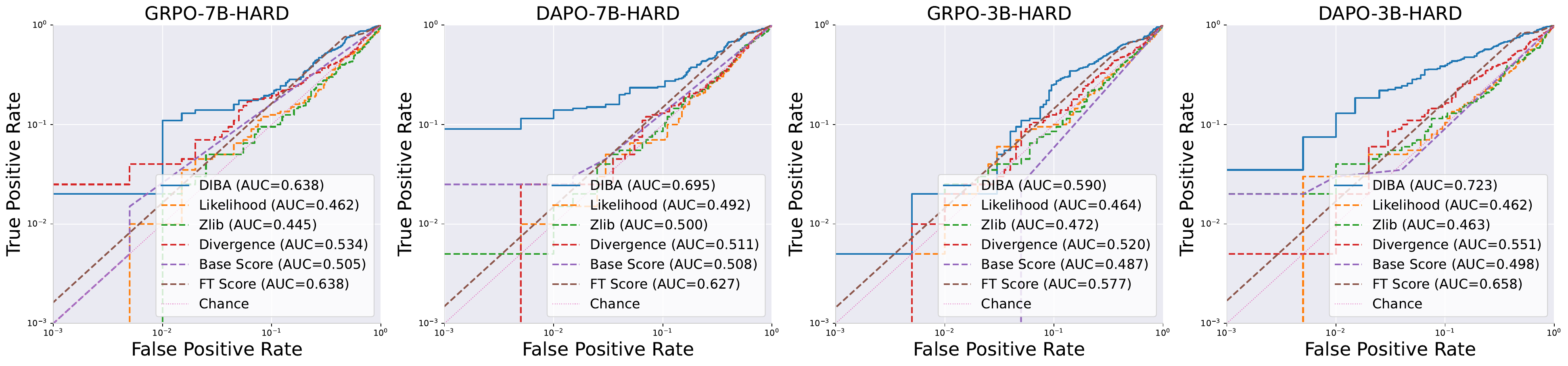}
    \caption{In-distribution results of shadow datasets}
    \label{fig:roc_shadow}
\end{figure*}

\mypara{In-distribution Performance}
The in-distribution performance of DIBA on the two shadow datasets is shown in \Cref{fig:roc_shadow}. We observe a significant performance gap between them: the easy dataset achieves near-perfect AUC and TPR@0.1\%FPR, while the hard dataset exhibits substantially lower detection performance. 
To understand this discrepancy, we analyze the results from the perspective of training dynamics and generalization. 
As shown in \Cref{tab:shadow-reward}, the difference in detectability correlates with the degree of generalization exhibited by the model during RLVR training. 
On the easy dataset, the model generalizes poorly, leading to higher performance improvement on member prompts than non-member ones. 
In contrast, the hard dataset generalizes well, resulting in a smaller reward gap. 
This reduces the reward gap between train and test sets, weakening the behavioral divergence signal and making membership harder to detect. 
Thus, membership distinguishability is not solely determined by training exposure, but by the extent to which learning fails to generalize.

\begin{table*}[]
\centering
\caption{Train and test reward before and after training. }
\label{tab:shadow-reward}
\resizebox{0.9\textwidth}{!}{\begin{tabular}{c|cccc|cccc} \toprule
\multirow{3}{*}{Model} & \multicolumn{4}{c|}{Easy} & \multicolumn{4}{c}{Hard} \\ \cmidrule{2-9} 
 & \multicolumn{2}{c|}{Begin} & \multicolumn{2}{c|}{End} & \multicolumn{2}{c|}{Begin} & \multicolumn{2}{c}{End} \\ \cmidrule{2-9} 
 & Train Reward & \multicolumn{1}{c|}{Test Reward} & Train Reward & Test Reward & Train Reward & \multicolumn{1}{c|}{Test Reward} & Train Reward & Test Reward \\ \midrule
GRPO-3B & 0.521 & \multicolumn{1}{c|}{0.406} & 0.779 & 0.459 & 0.490 & \multicolumn{1}{c|}{0.486} & 0.671 & 0.565 \\
DAPO-3B & 0.541 & \multicolumn{1}{c|}{0.379} & 0.882 & 0.449 & 0.495 & \multicolumn{1}{c|}{0.486} & 0.839 & 0.558 \\
GRPO-7B & 0.469 & \multicolumn{1}{c|}{0.437} & 0.893 & 0.565 & 0.482 & \multicolumn{1}{c|}{0.487} & 0.754 & 0.621 \\
DAPO-7B & 0.493 & \multicolumn{1}{c|}{0.396} & 0.938 & 0.563 & 0.481 & \multicolumn{1}{c|}{0.467} & 0.848 & 0.612 \\\bottomrule
\end{tabular}
}\end{table*}

\section{Adaptation to VLMs}\label{app:vlms}
\mypara{Training Setups}
We apply GRPO and DAPO for fine-tuning the VLM using rule-based rewards (e.g., answer correctness via symbolic verification). 
However, due to limited optimization of current RL frameworks (e.g., \texttt{verl}) for VLMs, training is significantly slower (40 hours for GRPO and 68 hours for DAPO) and less stable compared to language-only models.
Notably, a key component, the overlong response buffer, in DAPO is not yet supported for VLMs, limiting the full replication of advanced training dynamics.
Despite these challenges, we successfully complete RLVR training and observe measurable policy improvement on the task. 

\mypara{Implementation Details}
We sample 8 responses per prompt using temperature 0.5, top-p=0.9, and a repetition penalty of 1.05. 
All input images are resized to 448×448  and encoded using the built-in vision tower. 
The predictor is trained using the same stacking classifier as the previous setting.
\begin{table}[h]
\centering
\caption{Hard Example Analysis on VLMs.}
\label{tab:hard-vl}
\large
\resizebox{\linewidth}{!}{\begin{tabular}{c|c|ccc|ccc}\toprule
\multirow{2}{*}{Model} & \multirow{2}{*}{Split} & \multicolumn{3}{c|}{GRPO} & \multicolumn{3}{c}{DAPO} \\ \cmidrule{3-8} 
 &  & \# Samples & AUC & TPR@0.1\%FPR & \# Samples & AUC & TPR@0.1\%FPR \\ \midrule
\multirow{4}{*}{VL-7b} & All-0 & 111 & 0.527 & 0.046 & 115 & 0.492 & 0.032 \\
 & All-1 & 47 & 0.477 & 0.042 & 47 & 0.382 & 0.067 \\
 & Residual & 243 & 0.807 & 0.078 & 239 & 0.771 & 0.059 \\ \cmidrule{2-8} 
 & Unified & 400 & 0.728 & 0.050 & 400 & 0.691 & 0.034 \\\bottomrule
\end{tabular}
}\end{table}

\mypara{Analysis on Example Hardness}
We extend the example hardness analysis from \Cref{sec:hardness} to the VLM setting, with results presented in \Cref{tab:hard-vl}. 
The significant performance gap across splits confirms the validity of our hardness characterization: prompts that are consistently incorrect (All-0) or consistently correct (All-1) across training, i.e., those inducing minimal policy updates, remain challenging for membership inference. 
Both groups exhibit AUC scores close to 0.5, indicating near random-guess performance. 
This aligns with findings in the language-only setting and underscores that, even in multimodal RLVR, examples with little behavioral shift contribute weak leakage signals, making them inherently harder to distinguish as members. 
The results reinforce that membership distinguishability in VLMs, as in LLMs, is driven by measurable changes in model behavior rather than static input properties. 

\mypara{Fitting Level}
We also examine the training dynamics and fitting behavior of RLVR in VLMs. 
Compared to language-only LLMs, VLMs require significantly longer training epochs to achieve a good fit, often around 100 epochs under GRPO. 
This slower convergence reflects the added complexity of aligning visual inputs with reasoning trajectories. 
Notably, in DAPO, we observe negative average train rewards. 
This stems from the design of its reward function, which assigns explicit negative feedback to incorrect outputs rather than neutral (zero) rewards as in GRPO. 

Crucially, we find that exposure auditing performance remains strongly correlated with the reward gap (0.7 AUC on the training set and 0.46 on the test set), the difference between train and test reward. 
Larger gaps correspond to higher attack performance, consistent with our findings on LLMs. 
This confirms that, even in the multimodal setting, membership leakage in RLVR is driven by how much the model learns from its training data.

\begin{figure}[h]
    \centering
    \includegraphics[width=0.9\linewidth]{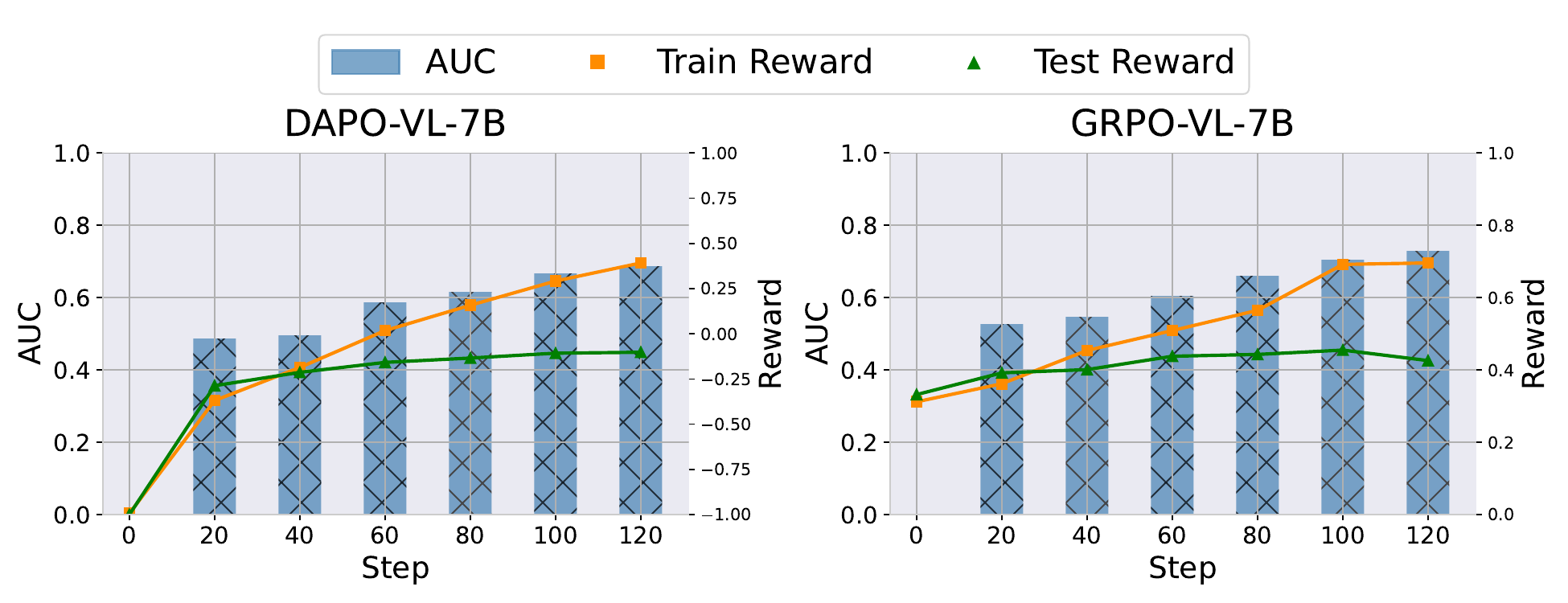}
    \caption{Performance over steps.}
    \label{fig:ablate_step_vl}
\end{figure}

\begin{figure}[h]
    \centering
    \includegraphics[width=1\linewidth]{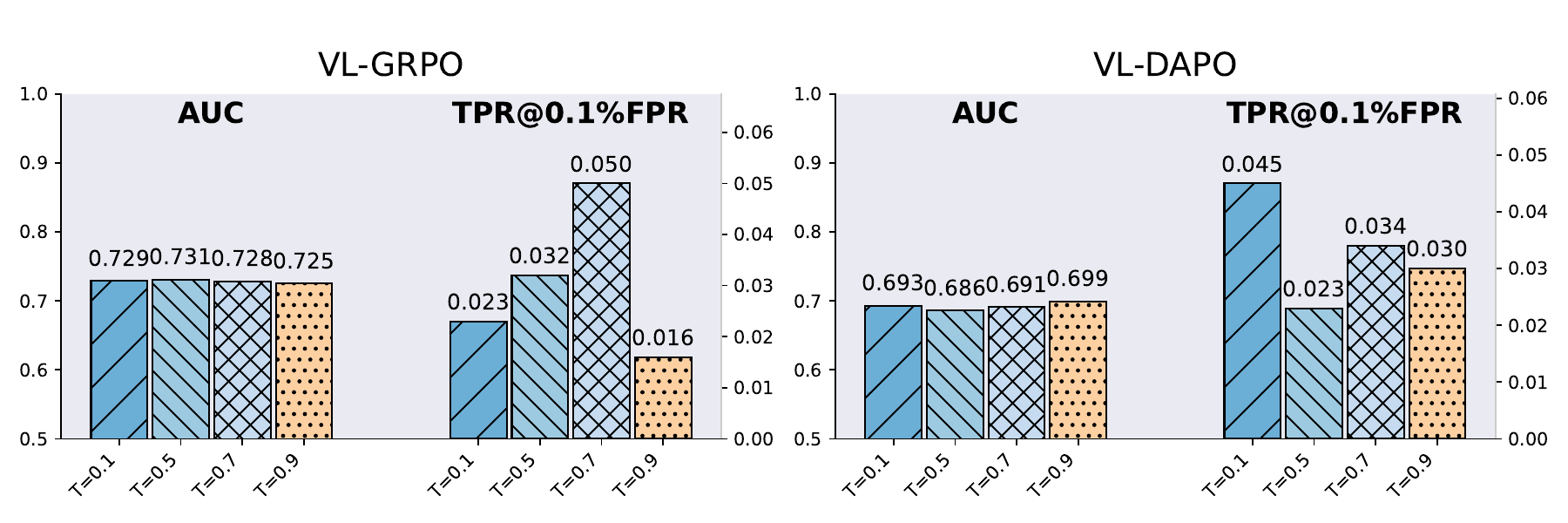}
        \includegraphics[width=1\linewidth]{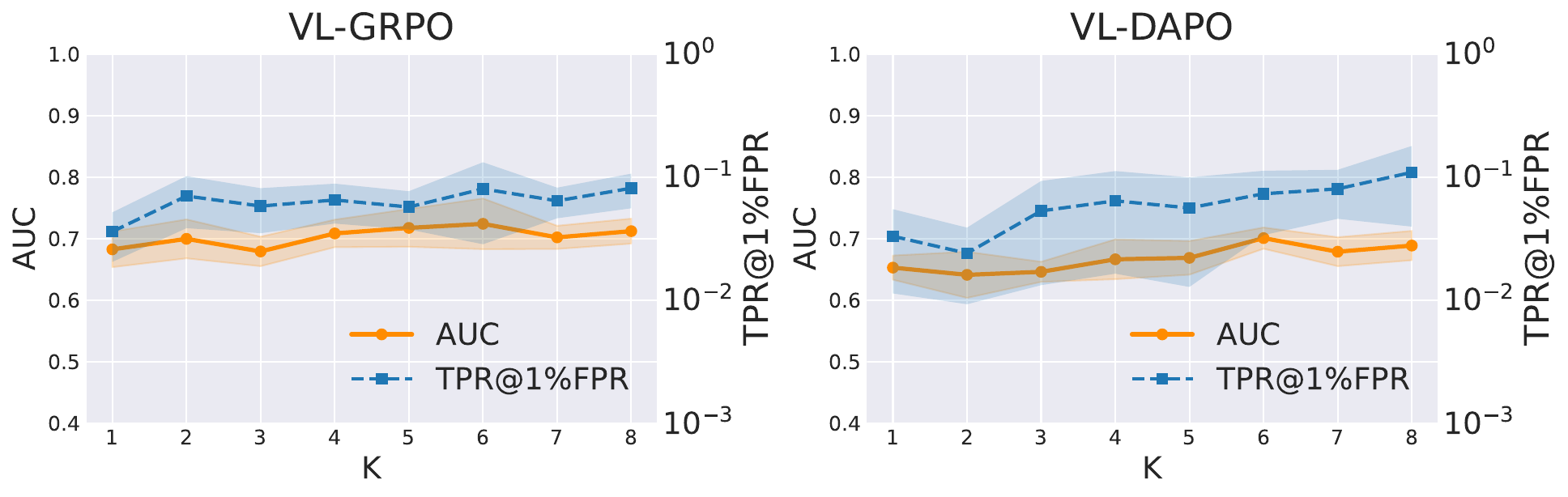}
    \caption{Ablation study on sampling temperature. }
    \label{fig:temp_ablate_vl}
\end{figure}

\mypara{Ablation on Temperature and Samples}
The ablation studies on sampling temperature and the number of generated responses for VLMs are presented in \Cref{fig:temp_ablate_vl}.
Regarding sampling temperature, we find that T=0.7  achieves the best overall trade-off, yielding high AUC and stable TPR@0.1\%FPR. 
While the DAPO model at T=0.1 exhibits an anomalously high TPR in one run, we treat this as noise: at such a low temperature, generation becomes nearly deterministic (akin to greedy decoding), with minimal response diversity. 
This limits reliable estimation of behavioral divergence and advantage signals, making the outlier result likely coincidental rather than indicative of improved detection. 

For the number of sampled responses, the trend mirrors our findings in the LLM setting: increasing k provides diminishing returns in AUC, but continues to improve TPR@0.1\%FPR significantly. 
This confirms that more samples better capture the tail behavior of the policy, enhancing precision in high-threshold regimes.  
Based on this analysis, we retain k=8  as the default number of responses for VLM evaluation, ensuring robust feature estimation without excessive computational cost.

\begin{table}[]
\centering
\caption{Accuracy (ACC) and TPR@0.1\%FPR for DIBA on VLMs.}
\label{tab:main-vl}
\resizebox{\linewidth}{!}{\begin{tabular}{c|c|cc|cc}\toprule
 &  & \multicolumn{2}{c|}{GRPO} & \multicolumn{2}{c}{DAPO} \\ \cmidrule{3-6} 
\multirow{-2}{*}{Model} & \multirow{-2}{*}{Metric} & ACC & {\color[HTML]{000000} TPR@0.1\%FPR} & ACC & {\color[HTML]{000000} TPR@0.1\%FPR} \\ \midrule
 & FT Score & 0.663 & 0.002 & 0.646 & 0.002 \\
 & Base Score & 0.474 & 0.002 & 0.474 & 0.001 \\
 & Divergence & 0.514 & 0.002 & 0.498 & 0.010 \\
 & Entropy & 0.477 & 0.013 & 0.513 & 0.003 \\
 & Likelihood & 0.516 & 0.013 & 0.508 & 0.003 \\ \cmidrule{2-6} 
\multirow{-6}{*}{VL-7b} & DIBA & 0.679 & 0.050 & 0.643 & 0.034 \\ \bottomrule
\end{tabular}
}\end{table}

\section{Defense}\label{app:defense}

\mypara{Perturbation Config}
We stochastically apply the following operations with the indicated probabilities within the perturbation budget.
\begin{itemize}[leftmargin=*]
    \item     Character swap (p=20\%): Randomly swap two adjacent characters (e.g., “hello” → “hlelo”).
    \item     Character deletion (p=10\%)): Randomly remove a character from the text (e.g., “text” → “tet”).
    \item     Keyboard-style typos (p=10\%)): Replace a character with one of its neighboring keys on a QWERTY keyboard (e.g., “cat” → “cay”).
    \item     Random casing (p=30\%)): Randomly toggle the case of individual letters (e.g., “Hello” → “hELlo”).
    \item     Punctuation insertion (p=10\%)): Insert a random punctuation mark at a random position (e.g., “hi” → “hi!”), modeling accidental keystrokes.
    \item     Punctuation removal (p=10\%)): Delete existing punctuation marks (e.g., “What?” → “What”).
    \item     Double whitespace injection (p=10\%)): Insert an extra space between words or tokens.
\begin{table}[b]
\centering
\caption{Defense Evaluation Against Perturbation: AUC and TPR@0.1\%FPR of DIBA against regularization. DAPO sets $\beta=0$ for default. Reward represents the accuracy on the test set.}
\label{tab:regularization}
\large
\resizebox{0.95\linewidth}{!}{
\begin{tabular}{c|c|ccc|ccc} \toprule
 &  & \multicolumn{3}{c|}{Math-3b (Reward 0.422)} & \multicolumn{3}{c}{Math-7b (Reward 0.430)} \\ \cmidrule{3-8} 
\multirow{-2}{*}{Model} & \multirow{-2}{*}{$\beta$} & AUC & { TPR@0.1\% FPR} & Reward & AUC & {\color[HTML]{000000} TPR@0.1\% FPR} & Reward \\ \midrule
 & 0.1 & 0.649 & 0.006 & 0.526 & 0.729 & 0.122 & 0.705 \\
 & 0.01 & 0.672 & 0.003 & 0.593 & 0.728 & 0.121 & 0.774 \\
\multirow{-3}{*}{GRPO} & 0.001 & 0.712 & 0.073 & 0.648 & 0.749 & 0.084 & 0.823 \\ \midrule
 & 0.1 & 0.835 & 0.151 & 0.812 & \multicolumn{3}{c}{} \\
 & 0.01 & 0.828 & 0.128 & 0.832 & \multicolumn{3}{c}{\multirow{-2}{*}{OOM}} \\
\multirow{-3}{*}{DAPO} & 0.0 & 0.834 & 0.154 & 0.846 & 0.818 & 0.107 & 0.911 \\\bottomrule
\end{tabular}
}\end{table}

\end{itemize}

\mypara{Local Differential Privacy}
We first establish the theoretical bound for the local DP setting.
\begin{definition}
(($\epsilon$, $\delta$)-DP\cite{dwork2006differential})
A randomized mechanism $\mathcal{A}$ satisfies $(\varepsilon,\delta)$-differential privacy if for any two neighboring inputs $\mathbf{f}$ and $\mathbf{f}'$ that differ in at most one element, and for any measurable output set $\mathcal{S}$,
\begin{equation}
\Pr[\mathcal{A}(\mathbf{f}) \in \mathcal{S}] \leq e^{\varepsilon} \cdot \Pr[\mathcal{A}(\mathbf{f}') \in \mathcal{S}] + \delta,
\end{equation}
where $\varepsilon > 0$ denotes the privacy budget, and $\delta$ is a relaxation parameter~\cite{dwork2006differential}.
\end{definition}
In practice, we adopt the Gaussian mechanism, defined as $\mathcal{A}(\mathbf{f}) = \mathbf{f} + \mathcal{N}(0, \sigma^2 \mathbf{I})$, which guarantees $(\varepsilon,\delta)$-DP, when the noise standard deviation $\sigma$ satisfies the condition derived in~\cite{wei2020federated}:
\begin{equation}
\sigma \geq \frac{\sqrt{2 \ln(1.25/\delta)} \cdot \Delta}{\varepsilon},
\end{equation}
where $\Delta = \max_{\mathbf{f} \sim \mathbf{f}'} \|\mathbf{f} - \mathbf{f}'\|_2$ is the $\ell_2$-sensitivity of the feature mapping.
Assume $\|\mathbf{f}\|_2 \le C$, we can get the $\ell_2$-sensitivity can be bounded as $\Delta \leq 2C$.
Therefore,
\begin{equation}
\varepsilon \geq \frac{2C \sqrt{2 \ln(1.25/\delta)}}{\sigma}.
\end{equation}
\begin{figure}
    \centering
    \includegraphics[width=0.9\linewidth]{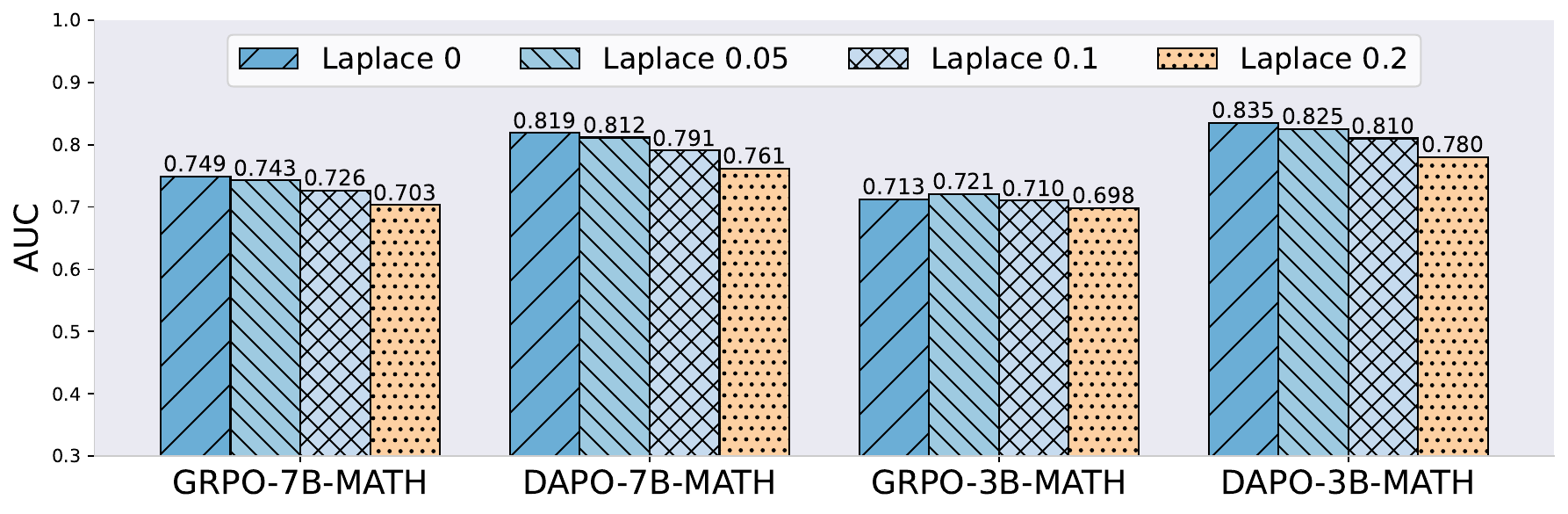}
    \includegraphics[width=0.9\linewidth]{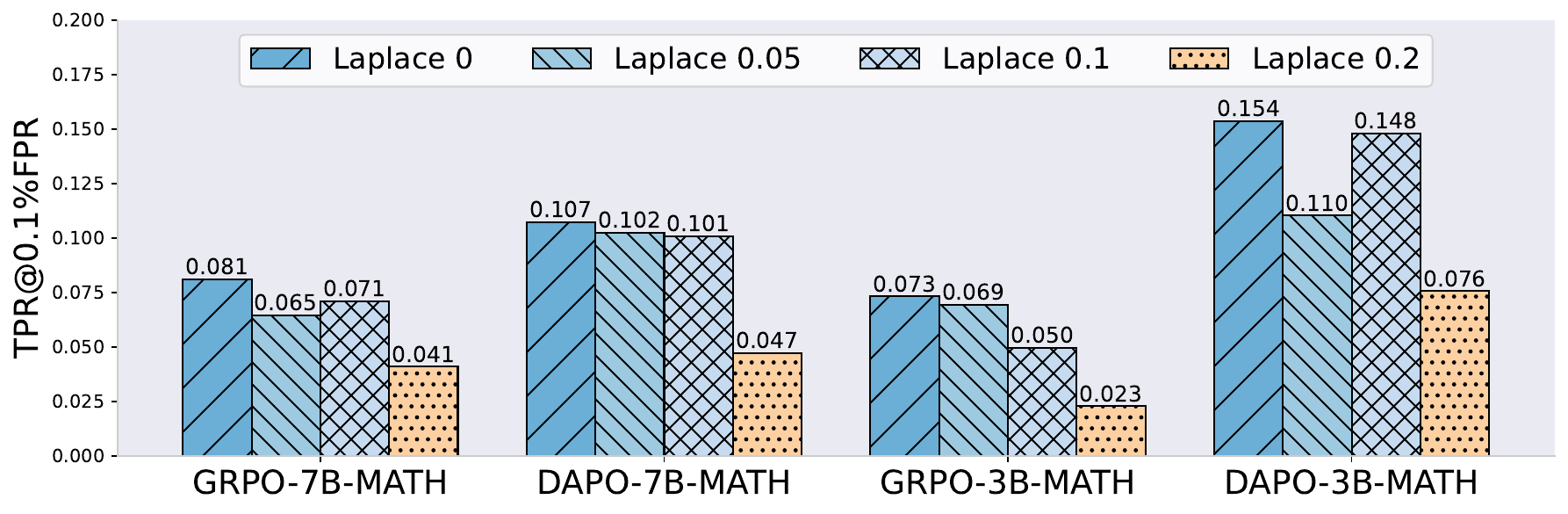}
    \caption{AUC and TPR@0.1\%FPR of DIBA against local DP with different Laplacian noises.}
    \label{fig:lap_noise}
\end{figure}

\begin{figure}[h]
    \centering
    \includegraphics[width=0.9\linewidth]{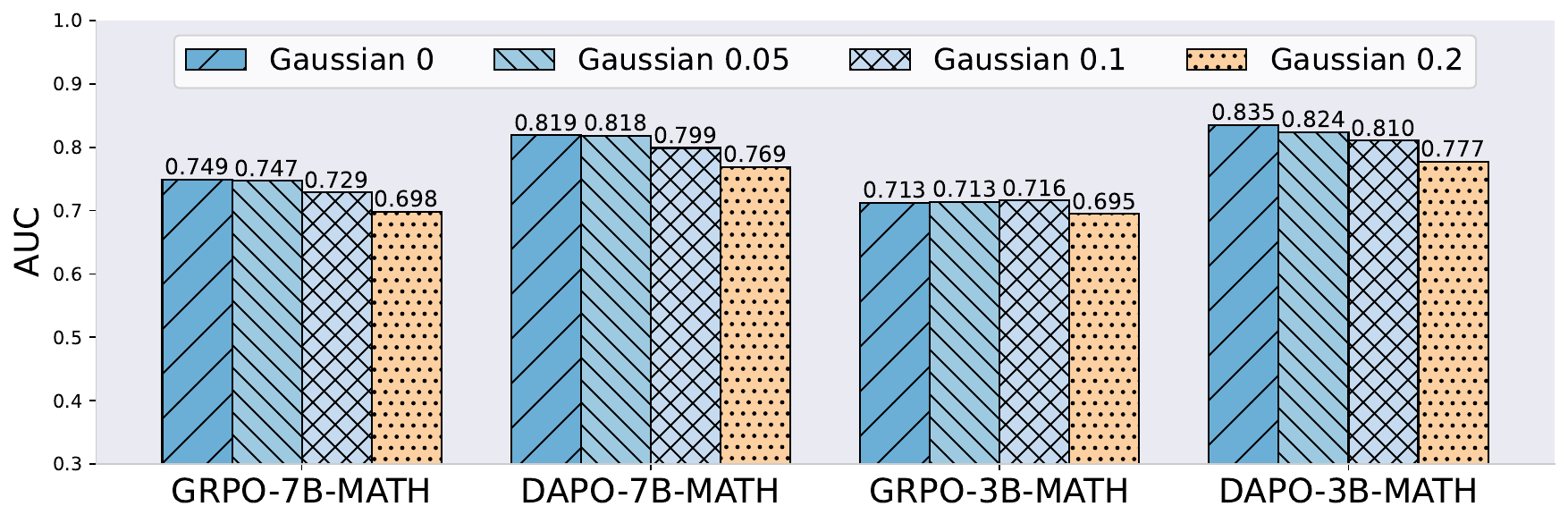}
    \includegraphics[width=0.9\linewidth]{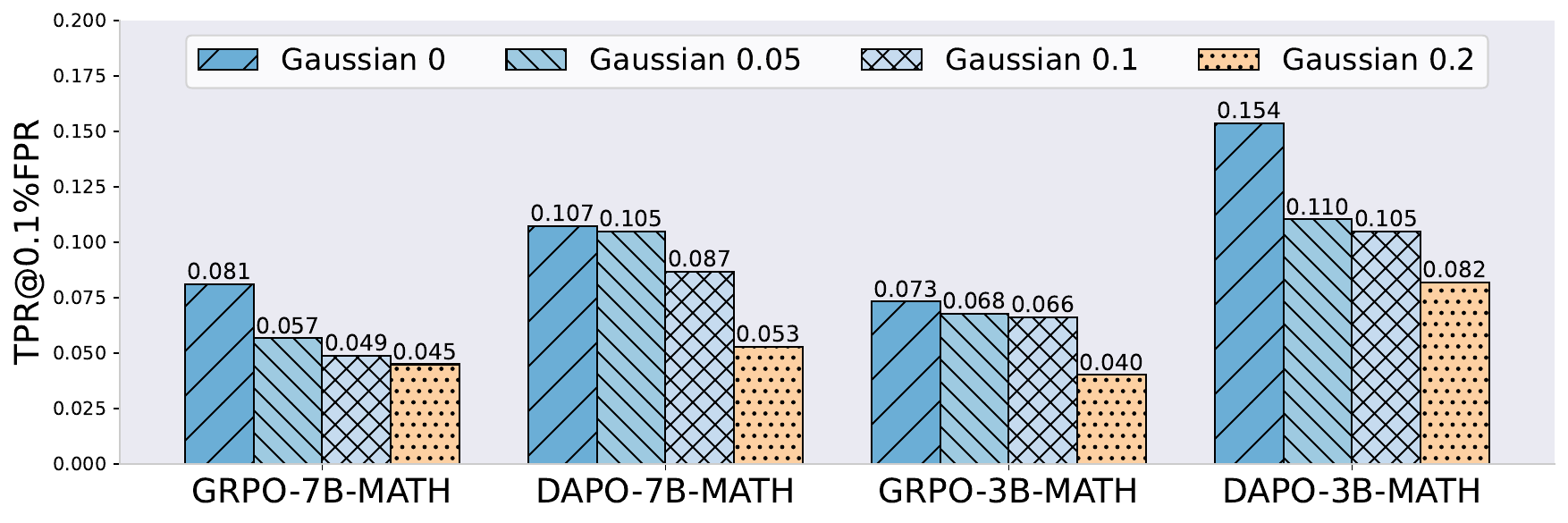}
    \caption{AUC and TPR@0.1\%FPR of DIBA against local DP with different Gaussian noise.
    }
    \label{fig:gau-auc}
\end{figure}

For the Laplace mechanism, it is defined as
$
\mathcal{A}(\mathbf{f}) = \mathbf{f} + \text{Lap}(\mathbf{0}, b),
$
which guarantees \(\varepsilon\)-differential privacy, where \(\text{Lap}(\mathbf{0}, b)\) denotes independent Laplace noise with scale \(b\) added to each feature dimension~\cite{dwork2006differential}. The noise scale \(b\) is determined according to the \(\ell_1\)-sensitivity of the feature mapping:
$
b \geq \frac{\Delta_1}{\varepsilon},
$
where 
$
\Delta_1 = \max_{\mathbf{f} \sim \mathbf{f}'} \|\mathbf{f} - \mathbf{f}'\|_1
$
is the \(\ell_1\)-sensitivity of the feature vector. 
Assuming that each feature vector satisfies \(\|\mathbf{f}\|_1 \le C_1\), the \(\ell_1\)-sensitivity can be bounded as \(\Delta_1 \le 2 C_1\). Therefore, the achievable privacy budget satisfies
$
\varepsilon \geq \frac{2 C_1}{b}.
$
We report the defense performance of adding Laplacian noise to the extracted feature in \Cref{fig:lap_noise}.

\begin{table}[b]
\centering
\caption{Performance in differentially private decoding.}
\label{tab:dp-decode}
\resizebox{0.95\linewidth}{!}{\begin{tabular}{c|c|cc|cc} \toprule
 &  & \multicolumn{2}{c|}{GRPO} & \multicolumn{2}{c}{DAPO} \\ \cmidrule{3-6} 
\multirow{-2}{*}{Model} & \multirow{-2}{*}{Strength} & AUC & {\color[HTML]{000000} TPR@0.1\%FPR} & AUC & {\color[HTML]{000000} TPR@0.1\%FPR} \\ \midrule
 & 0 & 0.712 & 0.073 & 0.834 & 0.154 \\
 & 0.01 & 0.715 & 0.034 & 0.837 & 0.111 \\
 & 0.05 & 0.714 & 0.040  & 0.837 & 0.171 \\
\multirow{-4}{*}{Math-3b} & 0.1 & 0.709 & 0.056 & 0.844 & 0.109 \\ \midrule
 & 0 & 0.749 & 0.084 & 0.818 & 0.107 \\
 & 0.01 & 0.758 & 0.082 & 0.803 & 0.100 \\
 & 0.05 & 0.759 & 0.048 & 0.822 & 0.073 \\
\multirow{-4}{*}{Math-7b} & 0.1 & 0.747 & 0.071 & 0.822 & 0.112 \\\bottomrule
\end{tabular}
}\end{table}

\section{Instruction Following Dataset}\label{app:if}
MulDimIF~\cite{ye2025multi} performs constraint expansion, conflict detection, and instruction rewriting based on the IFEval dataset~\cite{zhou2023instruction}, yielding a set of code-verifiable instruction-following test samples.
The training setting is the same as previous GRPO training, with epoch set to 2 and batch size set to 32.

\section{Tool Use Dataset}\label{app:tool}
We use the \texttt{xlam-function-calling-60k} dataset~\cite{liu2024apigen}, which provides structured function-calling samples with ground-truth tool invocations.
Each sample specifies available tools with typed parameter schemas; the model must produce correctly formatted \texttt{<tool\_call>} outputs with accurate function names and arguments.
The reward combines format compliance, argument correctness, and response length control.
Training uses GRPO on Qwen2.5-3B-Instruct for 2 epochs with a batch size of 32.
For the membership inference evaluation, we use correctness reward alone (binary verification of function name and parameter matching) as the behavioral score, and sample held-out prompts from the same dataset distribution as non-members.

\end{document}